\documentclass[10pt,onecolumn]{elsart5p}
\usepackage{graphicx}
\usepackage{dcolumn}
\usepackage{amsmath}
\usepackage{amssymb}
\usepackage{bm}
\usepackage{siunitx}
\usepackage{natbib}
\usepackage{color}

\newcommand{\elnino}{El Ni\~no}
\newcommand{\threshold}{\Delta T_\text{th}}
\newcommand{\degC}[1]{\SI{#1}{\degree C}}
\newcommand{\degN}[1]{\SI{#1}{\degree N}}
\newcommand{\degS}[1]{\SI{#1}{\degree S}}
\newcommand{\degE}[1]{\SI{#1}{\degree E}}

\newcommand{\Few}{F_\text{e-w}}
\newcommand{\Fopt}{F_\text{s}}
\newcommand{\figref}[1]{Fig.~\ref{fig:#1}}

\newcommand{\highlight}[1]{\color{black}#1}

\DeclareMathOperator{\sign}{sign}

\begin{document}

\begin{frontmatter}

\title{Leaders and followers: Quantifying consistency in spatio-temporal
propagation patterns}

\author[ISC]{Thomas Kreuz}\ead{thomas.kreuz@cnr.it},
\author[ISC,UF,VU]{Eero Satuvuori},
\author[UB]{Martin Pofahl},
\author[ISC]{Mario Mulansky}

\address[ISC]{Institute for complex systems, CNR, Sesto Fiorentino, Italy}
\address[UF]{Department of Physics and Astronomy, University of Florence,
Sesto Fiorentino, Italy}
\address[VU]{Move Research Institute Amsterdam, Department of Human Movement
Sciences, Vrije Universiteit Amsterdam, The Netherlands}
\address[UB]{Department of Epileptology, University of Bonn, Germany}

\date{\today}

\begin{abstract}
Repetitive spatio-temporal propagation patterns are encountered in fields as 
wide-ranging as climatology, social communication and network science. In 
neuroscience, perfectly consistent repetitions of the same global propagation
pattern are called a \textit{synfire pattern}.
For any recording of sequences of discrete events (in neuroscience terminology:
sets of spike trains) the questions arise how closely it resembles such a
synfire pattern and which are the spike trains that lead/follow.
Here we address these questions and introduce an algorithm built on two new
indicators, termed \textit{SPIKE-Order} and \textit{Spike Train Order}, that
define the \textit{Synfire Indicator} value, which allows to sort multiple
spike trains from leader to follower and to quantify the consistency of the
temporal leader-follower relationships for both the original and the optimized
sorting.
We demonstrate our new approach using artificially generated datasets before we
apply it to analyze the consistency of propagation patterns in two real
datasets from neuroscience (Giant Depolarized Potentials
in mice slices) and climatology (El Ni\~{n}o sea surface temperature
recordings).
The new algorithm is distinguished by conceptual and practical simplicity, low
computational cost, as well as flexibility and universality.
\end{abstract}

\end{frontmatter}


\newcommand{\abb}{\small\sf}

%
%

\section{\label{s:Intro} Introduction}

Recordings of spatio-temporal activity are ubiquitous in many scientific 
disciplines.
Among the most prominent examples are large-scale electrophysiological
measurements of neuronal firing patterns in experimental neuroscience
\citep{Spira13, Lewis15} and sensor data acquisition in seismology
\citep{Marano14}, oceanography \citep{Heidemann12}, meteorology
\citep{Muller13}, or climatology \citep{Menne12}.
Other examples include interaction protocols in social communication 
\citep{Lazer09, Gallos12} or monitoring single-node dynamics in network science 
\citep{Boccaletti14}.

In all of these fields recordings often exhibit well-defined patterns of 
spatio-temporal propagation where some prominent feature first appears at a 
specific location and then spreads to other areas until potentially becoming a 
global event.
Such characteristic propagation patterns occur in phenomena such as avalanches
\citep{Lacroix12}, tsunamis \citep{Pelinovsky06}, chemical waves and diffusion
processes \citep{Kuramoto12}, and epileptic seizures \citep{Baier12}.
Further examples are the epidemic transmission of diseases \citep{Belik11},
and, more recently, the spreading of memes on social networks \citep{Wei13} or
in science \citep{Kuhn14}.

\begin{figure}
    \includegraphics[width=85mm]{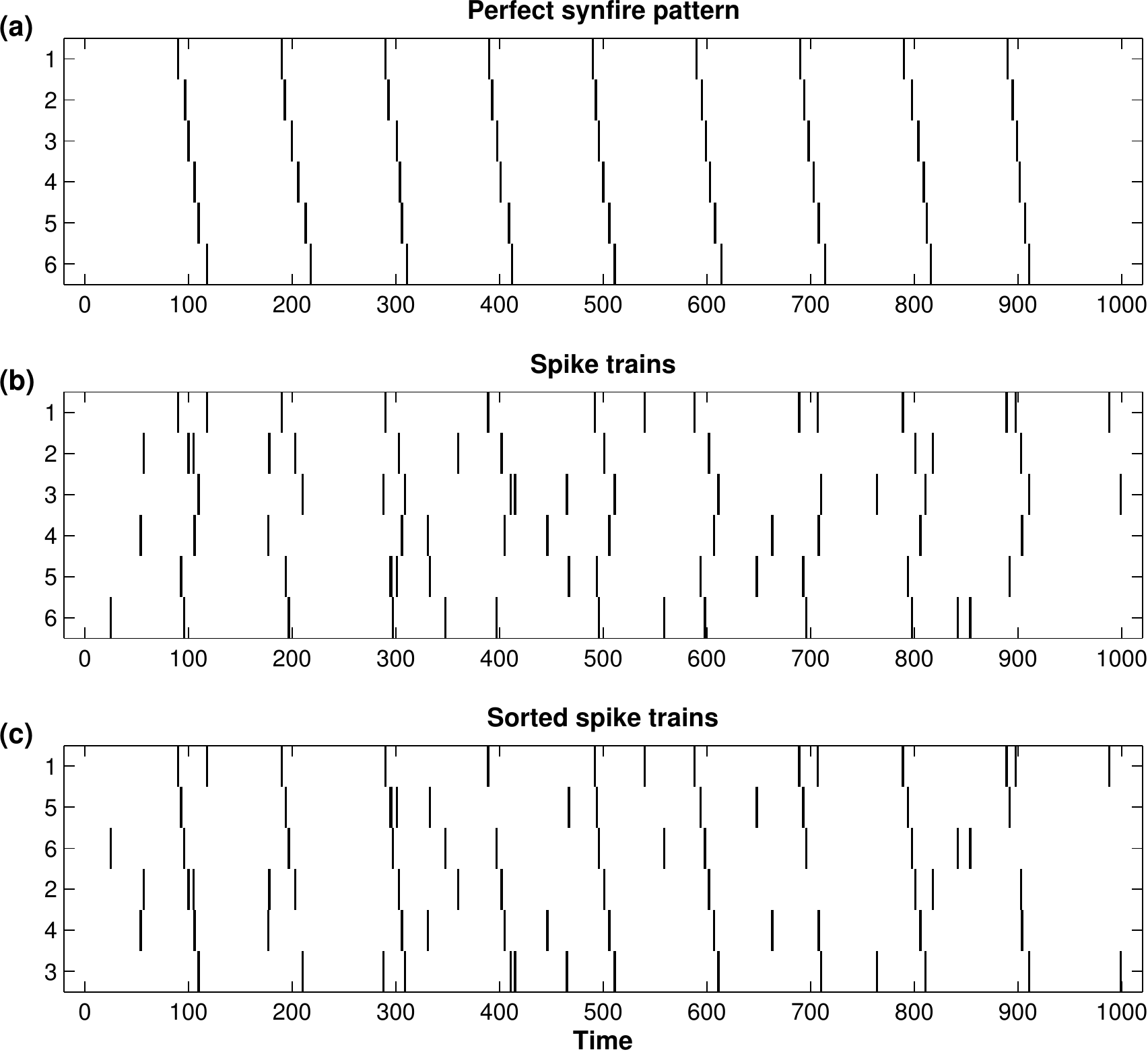}
    \caption{\abb\label{fig:SPIKE_Order_Motivation_Fig1} Motivation for 
SPIKE-order and Spike Train Order.   (a) Perfect synfire pattern.    (b) Unsorted 
set of spike trains.   (c) The same spike trains as in B but now sorted from leader to 
follower.}
\end{figure}

In many cases spatio-temporal recordings can be represented as a two-dimensional 
plot where for each recording site the occurrence of certain discrete events 
(often obtained from threshold crossings in continuous data) are indicated by 
time markers.
In neuroscience such a plot is known as a raster plot.
A sequence of stereotypical neuronal action potentials (\textit{spikes},
\cite{Rieke96}) is a \textit{spike train} and a set
of spike trains exhibiting perfectly consistent repetitions of the same global
propagation pattern is called a \textit{synfire pattern}.
In this paper we adapt this terminology and use all of these expressions not
only in the context of neuronal spikes but also for any other kind of discrete
events.
However, note that our use of the term `synfire pattern' differs slightly from
the literature (see e.g. \cite{Abeles09}).
Here we define a synfire pattern as a sequence of global events in which all
neurons fire in consistent order and the interval between successive events is
at least twice as large as the propagation time within an event.
An example of a rasterplot with spike trains forming a perfect synfire pattern
is shown in Fig. \ref{fig:SPIKE_Order_Motivation_Fig1}a.

For any spike train set exhibiting propagation patterns the questions arise
naturally whether these patterns show any consistency, i.e., to what extent do 
the spike trains resemble a synfire pattern, are there spike trains that 
consistently lead global events and are there other spike trains that 
invariably follow these leaders?
Such questions about leader-follower dynamics have been specifically
investigated not only in neuroscience \citep{Pereda05}, but also in fields as
wide-ranging as, e.g., climatology \citep{Boers14}, social communication
\citep{Varni10}, and human-robot interaction \citep{Rahbar15}.

%
%
%
%

\highlight{In this study we introduce a framework consisting of two directional measures
(\textit{SPIKE-Order} and \textit{Spike Train Order}) that allows to define a
value termed \textit{Synfire Indicator} which quantifies the consistency of the
leader-follower relationships in a rigorous and automated manner.}
This Synfire Indicator attains its maximal value of $1$ for a perfect synfire
pattern in which all neurons fire repeatedly in a consistent order from leader
to follower (Fig. \ref{fig:SPIKE_Order_Motivation_Fig1}a).

The same framework also allows to sort multiple spike trains from leader to
follower, as illustrated in Figs. \ref{fig:SPIKE_Order_Motivation_Fig1}b and
\ref{fig:SPIKE_Order_Motivation_Fig1}c.
This is meant purely in the sense of temporal sequence. Whereas Fig.
\ref{fig:SPIKE_Order_Motivation_Fig1}b shows an artificially created but rather
realistic spike train set, in Fig. \ref{fig:SPIKE_Order_Motivation_Fig1}c the
same spike trains have been sorted to become as close as possible to a synfire
pattern.
Now the spike trains that tend to fire first are on top whereas spike trains
with predominantly trailing spikes are at the bottom. 

We demonstrate the new approach using artificially generated datasets before we 
apply it to analyze the consistency of propagation patterns in two real datasets 
from very different fields of research, neuroscience and climatology.
The neurophysiological dataset consists of neuronal activity recorded from mice 
brain slices.
These recordings typically exhibit a sequence of global events termed Giant
Depolarized Potentials (GDPs) and one of the main questions we investigate is
whether it is possible to identify neurons that consistently lead these events
(potential hub neurons, see \cite{Bonifazi09}).
The climate dataset uses Optimum Interpolated Sea Surface Temperature (OISST)
data provided by the National Oceanic and Atmospheric Administration (NOAA) to
follow the El Ni\~{n}o phenomenon in the central pacific region
\cite{Santoso13} over the range of $35$ years,
from 1982 to 2016.
We employ a threshold criterion to track the El Ni\~{n}o events and then quantify
the consistency of the longitudinal movement of the propagation front.

The remainder of the article is organized as follows:
In the Methods (Section \ref{s:Measures}) we first describe the coincidence
detection (Section \ref{ss:Coincidence-Detection}) and the symmetric measure
SPIKE-Synchronization (Section \ref{ss:SPIKE-Synchronization}).
Subsequently, we introduce the new directionality approach consisting of the
two measures SPIKE-Order and Spike Train Order (Section \ref{ss:SPIKE-Order})
as well as the Synfire Indicator (Section \ref{ss:Synfire-Indicator}) before
we discuss the use of SPIKE-Order surrogates to evaluate the statistical
significance of the results in Section \ref{ss:Statistical-Significance}.
The Results Section \ref{s:Results} consists of three Subsections detailing
applications of the new approach to artificially generated datasets (Section
\ref{ss:Results-Sim}), neurophysiological data (Section \ref{ss:Results-Neuro})
and climate data from the El Ni\~{n}o phenomenon (Section \ref{ss:Results-Climate}). 
Conclusions are drawn in Section \ref{s:Discussion}.
Finally, both real datasets are described in the Appendix, the
electrophysiological recordings in Appendix \ref{App-s:HM-Data}, and the El
Ni\~{n}o dataset) in Appendix \ref{App-s:ElNino-Data}.

%
%
\section{\label{s:Measures} Measures}

Analyzing leader-follower relationships in a spike train set requires a
criterion that determines which spikes should be compared against each other.
What is needed is a match maker, a method which pairs spikes in such a way
that each spike is matched with at most one spike in each of the other spike
trains.
This match maker already exists.
It is the adaptive coincidence detection first used as the fundamental
ingredient for the bivariate measure \textit{event synchronization}
\citep{QuianQuiroga02b, Kreuz07a}.

Event synchronization itself is symmetric and quantifies the overall level of 
synchrony from the number of quasi-simultaneous appearances of spikes.
It was proposed along with an asymmetric measure termed \textit{delay asymmetry}
which evaluates the temporal order among matching spikes in the two spike trains.

However, unfortunately both event synchronization and delay asymmetry are
defined for the bivariate case of two spike trains only, rely on sampled time
profiles, and have a very non-intuitive normalization.
For the symmetric variant we have already addressed these issues by proposing
\textit{SPIKE-Synchronization} \citep{Kreuz15}, a renormalized multivariate
extension of event synchronization.

The two new measures SPIKE-Order and Spike Train Order proposed here improve
and extend the asymmetric measure delay asymmetry in the same way.
In particular, instead of just quantifying bivariate directionality they open
up a completely new application, since they allow us to sort the spike trains
according to the typical relative order of their spikes and to quantify the
consistency of this order using the Synfire Indicator.

All four approaches (bivariate/multivariate, symmetric/asymmetric) are
time-resolved as well as parameter- and scale-free.
Their calculation consists of two steps, adaptive coincidence detection
followed by a combination of normalization and windowing.
The first step, adaptive coincidence detection, is the same for all of these
measures.

\subsection{\label{ss:Coincidence-Detection} Adaptive Coincidence Detection}

Most coincidence detectors rely on a coincidence window of fixed size $\tau$ 
\citep{Kistler97, Naud11}.
However, since in many cases it is very difficult to judge whether two spikes
are coincident or not without taking the  local context into account (see
Fig. \ref{fig:SPIKE_Order_Coincidence-Detection_comb}a 
for an example), \citet{QuianQuiroga02b} proposed a more flexible 
coincidence detection.
This coincidence detection is scale- and thus parameter-free since the minimum
time lag $\tau^{(1,2)}_{ij}$ at which two spikes $t_i^{(1)}$ and $t_j^{(2)}$ of spike
trains $(1)$ and $(2)$ are no longer considered to be synchronous is adapted to the
local firing rates according to 
\begin{equation} \label{eq:Coincidence-MaxDist}
 \begin{aligned}
    \tau^{(1,2)}_{ij} = \min \{&t_{i+1}^{(1)} - t_i^{(1)}, t_i^{(1)} - t_{i-1}^{(1)},\\
    &t_{j+1}^{(2)} - t_j^{(2)}, t_j^{(2)} - t_{j-1}^{(2)}\}/2.
 \end{aligned}
\end{equation}

For some applications it might be appropriate to additionally introduce a 
maximum coincidence window $\tau_{max}$ as a parameter.
This way additional knowledge about the data (such as typical propagation
speed) can be taken into account in order to guarantee that two coincident
spikes are really part of the same propagation front.
Such a maximum coincidence window will be used in the application to the El
Ni\~{n}o climate data in Section \ref{ss:Results-Climate}.

\subsection{\label{ss:SPIKE-Synchronization} SPIKE-Synchronization}

In normalization and windowing SPIKE-Synchronization \citep{Kreuz15} has
evolved so substantially from event synchronization that here we refrain from
going into any detail on the original measure, but rather just mention the
main improvements.
For a thorough introduction to event synchronization please refer to the
original paper \citep{QuianQuiroga02b}, a more detailed comparison of the two
measures can be found in \citet{Kreuz15}.

The main difference is that SPIKE-Synchronization \citep{Kreuz15} results in a
discrete, not a continuous, spike-timing based profile.
The coincidence criterion is quantified by means of a coincidence indicator
\begin{equation} \label{eq:Bi12-SPIKE-Synchronization}
	C_i^{(1,2)}=\begin{cases}
		1 & {\rm if}  \min_j (|t_i^{(1)} - t_j^{(2)}|) < 
\tau_{ij}^{(1,2)} \cr
		0 & {\rm otherwise}
	\end{cases}
\end{equation}
which assigns to each spike either a one or a zero depending on whether this
spike is part of a coincidence or not.
Note that here, unlike for event synchronization, the minimum function and the
'$<$' guarantee that a spike can at most be coincident with one spike (the
nearest one) in the other spike train.
In case a spike is right in the middle between two spikes from the other spike
train there is no ambiguity since this spike is not coincident with either one
of them.

%
%
\begin{figure}
    \includegraphics[width=85mm]{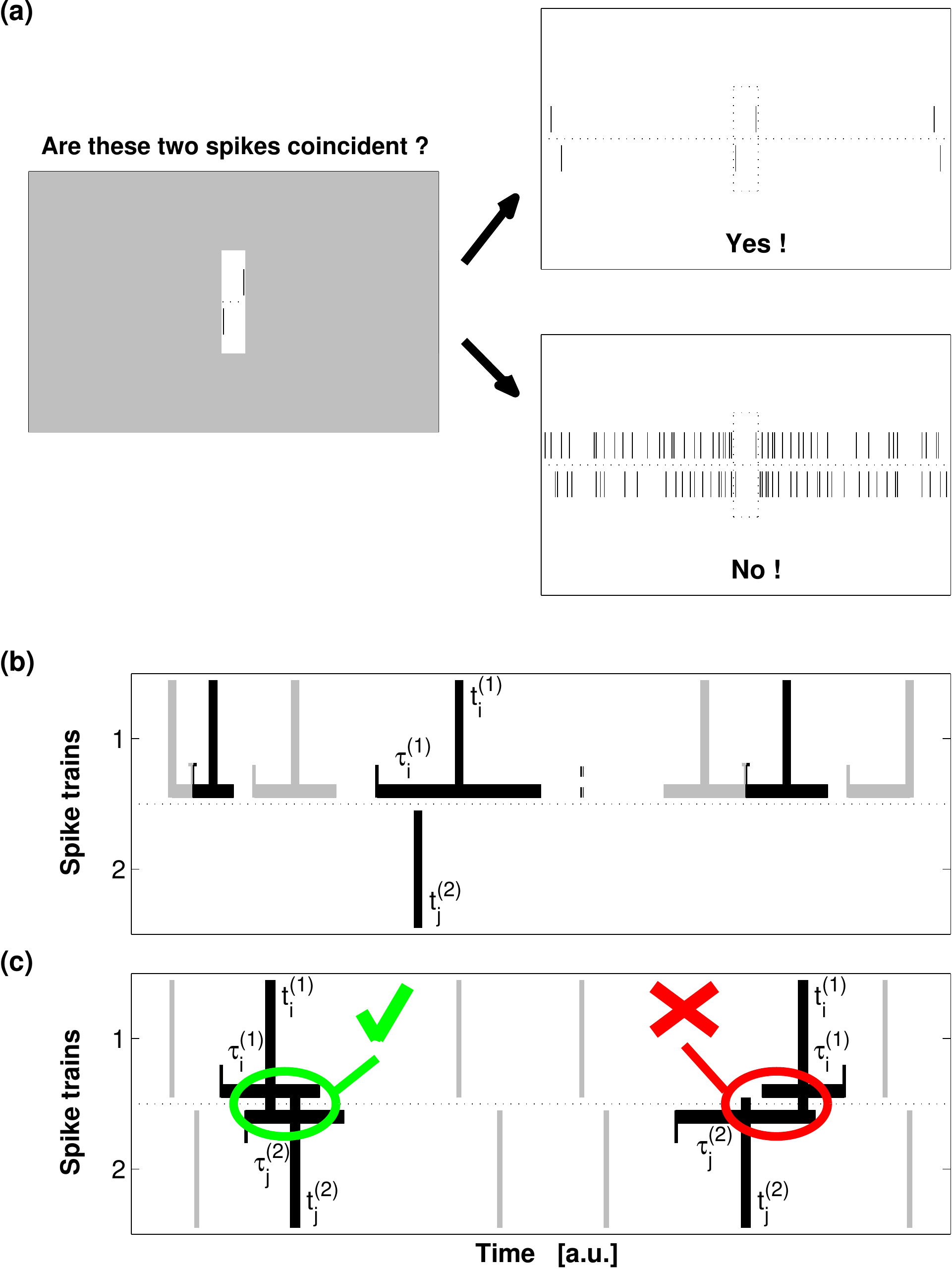}
    \caption{\abb\label{fig:SPIKE_Order_Coincidence-Detection_comb} (a) This
example demonstrates the usefulness of an adaptive coincidence detection.
Depending on  context the same two spikes (left) can appear as coincident
(right, top) or as  non-coincident (right, bottom). 
(b) Illustration of the adaptive coincidence detection.
For clarity spikes and their coincidence  windows are shown alternatively in
bright and dark color.
The first step assigns to each spike $t_i^{(1)}$ of the first spike train a
potential coincidence window which does not overlap with any other coincidence
window:
$\tau_i^{(1)} = \min \{t_{i+1}^{(1)} - t_i^{(1)}, t_i^{(1)} - t_{i-1}^{(1)}\}/2$.
Thus any spike from the second spike train can at most be coincident with one
spike from the first spike train.
Small vertical lines mark the times right in the middle between two spikes,
and a line is dashed when it does not mark the edge of a coincidence window.
(c) In the same way a coincidence window
$\tau_j^{(2)} = \min \{t_{j+1}^{(2)} - t_j^{(2)}, t_j^{(2)} - t_{j-1}^{(2)}\}/2$
is defined for spike $t_j^{(2)}$ from the second spike train.
For two spikes to be coincident they both have to lie in each other's
coincidence window which means that their absolute time difference has to be
smaller than $\tau_{ij}=\min \{\tau_i^{(1)}, \tau_j^{(2)}\}$ (which is
equivalent to the shorter definition found in Eq. \ref{eq:Coincidence-MaxDist}).
For the two spikes $t_i^{(1)}$ and $t_j^{(2)}$ on the left side this is the
case, whereas the spikes on the right side are not coincident.}
\end{figure}
%
%
This unambiguity, illustrated in Fig.
\ref{fig:SPIKE_Order_Coincidence-Detection_comb}b, is the essential property
which allows the adaptive coincidence detection to act as a match-maker for the
subsequent application of SPIKE-Synchronization.
Fig. \ref{fig:SPIKE_Order_Coincidence-Detection_comb}c shows examples, one with two
coincident and one with two non-coincident spikes.
%
%
\begin{figure}
    \includegraphics[width=85mm]{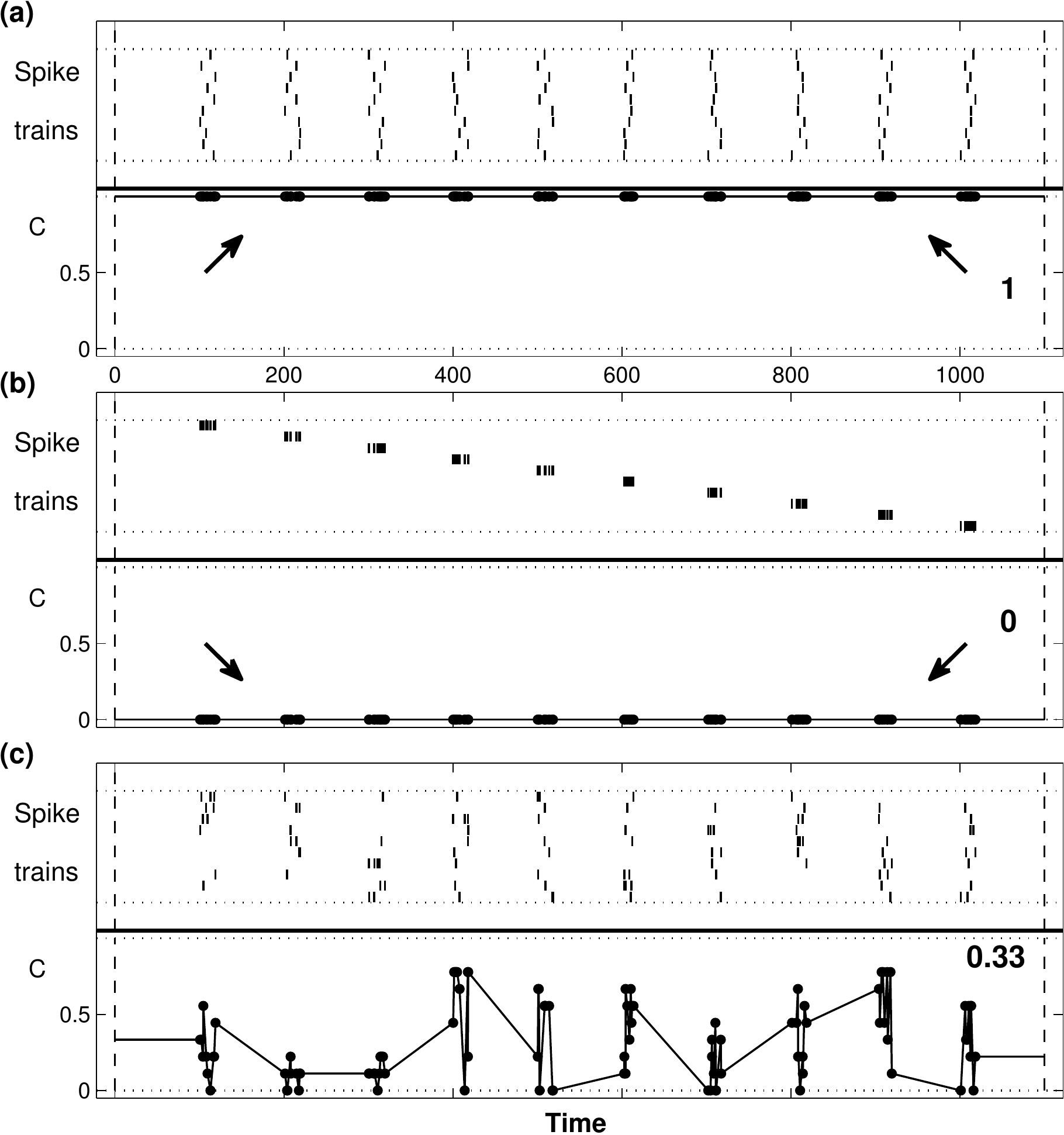}
    \caption{\abb\label{fig:SPIKE_Order_SPIKE_Sync_3}
SPIKE-Synchronization.
Note that the profile $C (t_k)$ is defined only at the times of the spikes but
a better visualization is achieved by connecting the individual dots. 
By construction the pooled spike train of these examples is identical consisting 
of $10$ evenly spaced bursts.
The only difference is the distribution of the spikes among the individual spike
trains which varies from maximum to minimum via intermediate synchrony.
SPIKE-Synchronization correctly indicates these changes.
(a) Maximum reliability results in the value one over the whole time interval.
Each spike train contains one spike per firing event.
(b) Synfire pattern of bursts resulting in minimum reliability corresponding to
the value zero for the whole time interval.
(c) A random distribution of spikes among spike trains yields intermediate values.}
\end{figure}
%
%

A multivariate version of SPIKE-Synchronization can be defined by generalizing
the bivariate coincidence detection of Eq. \ref{eq:Bi12-SPIKE-Synchronization}
to all pairs of spike trains $(n,m)$ with $n,m=1,...,N$ and $N$ denoting the
number of spike trains:
\begin{equation} \label{eq:Bi-SPIKE-Synchronization}
	C_i^{(n,m)}=\begin{cases}
		1 & {\rm if}  \min_j (|t_i^{(n)} - t_j^{(m)}|) < 
\tau_{ij}^{(n,m)} \cr
		0 & {\rm otherwise.}
	\end{cases}
\end{equation}
Here $\tau^{(n,m)}_{ij}$ is defined equivalent to Eq. \ref{eq:Coincidence-MaxDist}.
Subsequently, for each spike of every spike train a normalized coincidence 
counter
\begin{equation} \label{eq:Multi-SPIKE-Synchronization-Counter}
	C_i^{(n)}=\frac{1}{N-1} \sum_{m \neq n} C_i^{(n,m)}
\end{equation}
is obtained by averaging over all $N-1$ bivariate coincidence indicators 
involving the spike train $n$.

In order to obtain a single multivariate similarity profile we pool the spikes
of all the spike trains as well as their coincidence counters:
\begin{equation} \label{eq:Multi-Profile}
    	\{C_k\} = \bigcup_n \{C_{i(k)}^{(n(k))} \},
\end{equation}
\highlight{where we map the spike train indices $n$ and the spike indices $i$
into a global spike index $k$ denoted by the mapping $i(k)$ and $n(k)$.}

Note that in case there exist perfectly coincident spikes, $k$ counts over all
of these spikes.
From this discrete set of coincidence counters $C_k$ the SPIKE-Synchronization
profile $C(t_k)$ is obtained via $C(t_k) = C_k$. 
Finally, SPIKE-Synchronization is defined as the average value of this profile
\begin{equation} \label{eq:Multi-SPIKE-Synchronization-Profile}
	S_C = \begin{cases}
		\frac{1}{M} \sum_{k=1}^M C(t_k) & {\rm if} \  M > 0 \cr
		\ \ \ \ \ \ \ \ \ 1 & {\rm otherwise}
	\end{cases}
\end{equation}
with $M=\sum_{n=1}^N M^{(n)}$ denoting the total number of spikes in the pooled 
spike train. 

This way we have used the same consistent framework for both the bivariate and
the multivariate case.
The former is just a special case of the latter.
The interpretation is very intuitive: SPIKE-Synchroniza\-tion quantifies the 
overall fraction of coincidences.
It reaches one if and only if each spike in every spike train has one matching
spike in all the other spike trains (or if there are no spikes at all), and it
attains the value zero if and only if the spike trains do not contain any
coincidences.
Examples for both of these extreme cases can be found in Fig. 
\ref{fig:SPIKE_Order_SPIKE_Sync_3}a and \ref{fig:SPIKE_Order_SPIKE_Sync_3}b and 
one intermediate example (random distribution of spikes among spike trains) is 
shown in Fig. \ref{fig:SPIKE_Order_SPIKE_Sync_3}c.
For a derivation of the expectation value for Poisson spike trains please refer
to \cite{Mulansky15}.

In the multivariate analysis proposed in this paper, SPIKE-Synchronization can
be used to filter the input to the algorithm.
In order to focus on propagation
patterns within truly global events it is possible to set a threshold value
$C_{thr}$ for the SPIKE-Synchronization profile $C(t_k)$.
This way only spikes with a coincidence value higher than this parameter
$C_{thr}$ are taken into account, all the other noisy background spikes are
simply ignored.
This kind of filter will be used in the analysis of the neurophysiological
datasets in Section \ref{ss:Results-Neuro}.

\subsection{\label{ss:SPIKE-Order} SPIKE-Order and Spike Train Order}

SPIKE-Synchronization assigns to each spike of a given spike train pair a 
bivariate coincidence indicator.
These coincidence indicators $C_i^{(n,m)}$, which are either $0$ or $1$,
are then averaged over spike train pairs and converted into one overall
profile $C(t_k)$ normalized between $0$ and $1$.
In exactly the same manner SPIKE-Order and Spike Train Order assign bivariate
order indicators to spikes.
Also these two order indicators, the asymmetric $D_i^{(n,m)}$ and the symmetric
$E_i^{(n,m)}$, which both can take the values $-1$, $0$, or $+1$, are averaged
over spike train pairs and converted into two overall profiles $D(t_k)$ and
$E(t_k)$ which are normalized between $-1$ and $1$.
The SPIKE-Order profile $D(t_k)$ distinguishes leading and following spikes,
whereas the Spike Train Order profile $E(t_k)$ provides information about the
order of spike trains, i.e. it allows to sort spike trains from leaders to
followers. 

%
%
\begin{figure}
    \includegraphics[width=85mm]{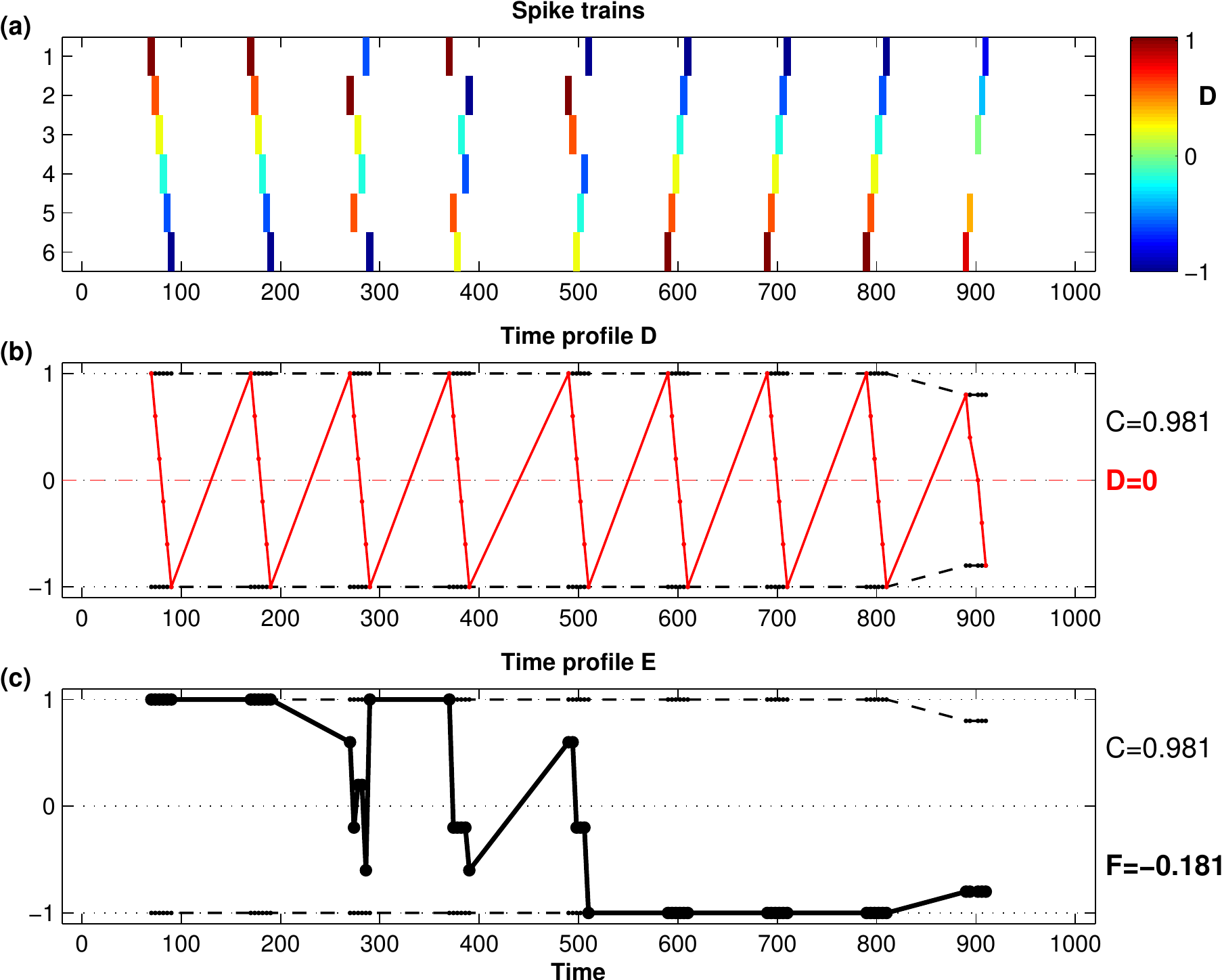}
    \caption{\abb\label{fig:SPIKE_Order_80_Standard_Example_1_Profiles} 
SPIKE-Order profile $D(t_k)$ and Spike Train Order profile $E(t_k)$ for an 
artificially created example dataset.
(a) The rasterplot shows $6$ spike trains which emit spikes in nine reliable
events.
For the first two events spikes fire in order, for the next three events the
order is random whereas for the last four events the order is inverted.
In the last event there is one spike missing.
Spike thickness decodes the SPIKE-Synchronization value $C(t_k)$ (here almost
constant), spike color the SPIKE-Order value $D(t_k)$.
(b,c) The SPIKE-Synchronization profile $C(t_k)$ and its mirror profile (dashed
black lines) act as envelope for both the SPIKE-Order profile $D(t_k)$ (b, red)
and the Spike Train Order profile $E(t_k)$ (c, black).
(b) The SPIKE-Order profile can not distinguish events with different firing 
order and by construction the average value is always $D = 0$.
(c) On the other hand, in the Spike Train Order profile events with different
firing order can clearly be distinguished.
For the first two correctly ordered events the value $1$ is obtained.
The next three events exhibit random order and correspondingly the profile
fluctuates rather wildly. 
Finally, the last four inversely ordered yield the value $-1$ except for the
last event for which the absolute minimum value can not be obtained since
one spike is missing.
The average value, the Synfire Indicator $F$, is not $0$ but negative which
reflects the dominance of the inversely ordered events.
}
\end{figure}
%
%

First of all, similar to the transition from the symmetric event synchronization 
to delay asymmetry, the symmetric coincidence indicator $C^{(n,m)}_i$ of 
SPIKE-Synchronization (Eq. \ref{eq:Bi-SPIKE-Synchronization}) is replaced by the 
asymmetric SPIKE-Order indicator 
\highlight{\begin{equation} \label{eq:Bi-SPIKE-Order-1}
	D_i^{(n,m)} = C_i^{(n,m)} \cdot \sign (t_{j'}^{(m)} - t_i^{(n)}),
\end{equation}}
\highlight{where the index $j'$ is defined from the minimum in Eq.
\ref{eq:Bi12-SPIKE-Synchronization} as $j' = \arg \min_j ( |t_i^{(1)} - t_j^{(2)}| )$.}

The corresponding value \highlight{$D_{j'}^{(m,n)}$} is obtained in an antisymmetric
manner as
\highlight{\begin{equation} \label{eq:Bi-SPIKE-Order-2}
	D_{j'}^{(m,n)} = C_{j'}^{(m,n)} \cdot \sign (t_i^{(n)} - t_{j'}^{(m)})
		    = - D_i^{(n,m)}.
\end{equation}}

Therefore, this indicator assigns to each spike either a $1$ or a $-1$ depending 
on whether the respective spike is leading or following a coincident spike from 
the other spike train.
The value $0$ is obtained for cases in which there is no 
coincident spike in the other spike train ($C^{(n,m)}_i = 0$), but also in cases 
in which the times of the two coincident spikes are absolutely identical  
\highlight{($t_{j'}^{(m)} = t_i^{(n)}$)}.

The multivariate profile $D(t_k)$ obtained analogously to Eq. 
\ref{eq:Multi-Profile} is normalized between $1$ and $-1$ and the extreme values 
are obtained if a spike is either leading ($+1$) or following ($-1$) coincident 
spikes in all other spike trains. It can be $0$ either if a spike is not part 
of any coincidences or if it leads exactly as many spikes from other spike 
trains in coincidences as it follows. From the definition in Eqs. 
\ref{eq:Bi-SPIKE-Order-1} and \ref{eq:Bi-SPIKE-Order-2} it follows immediately 
that $C_k$ is an upper bound for the absolute value $|D_k|$.

While the SPIKE-Order profile can be very useful for color-coding and
visualizing local spike leaders and followers (Fig. 
\ref{fig:SPIKE_Order_80_Standard_Example_1_Profiles}a), it is not useful as an 
overall indicator of Spike Train Order (Fig. 
\ref{fig:SPIKE_Order_80_Standard_Example_1_Profiles}b).
The profile is invariant under exchange of spike trains, i.e. it looks the same
for all events no matter what the order of the firing is (in our example only
the last event looks slightly different since one spike is missing).
Moreover, summing over all profile values, which is equivalent to summing over
all coincidences, necessarily leads to an average value of $0$, since for every
leading spike ($+1$) there has to be a following spike ($-1$). 

So in order to quantify any kind of leader-follower information between spike
trains we need a second kind of order indicator.
The Spike Train Order indicator is similar to the SPIKE-Order indicator defined
in Eqs. \ref{eq:Bi-SPIKE-Order-1} and \ref{eq:Bi-SPIKE-Order-2} but with two
important differences.
Both spikes are assigned the same value and this value now depends on the
order of the spike trains:
\highlight{
\begin{equation}  \label{eq:Bi-Spike-train-order-1}
 E_i^{(n,m)} = C_i^{(n,m)} \cdot
 				\begin{cases}
 					\sign (t_{j'}^{(m)} - t_i^{(n)})\quad\text{if}\quad n<m\\
                		\sign (t_i^{(n)} - t_{j'}^{(m)})\quad\text{if}\quad n>m
              	\end{cases}
\end{equation}}
and
\highlight{\begin{equation} \label{eq:Bi-Spik-train-order-2}
	E_{j'}^{(m,n)} = E_i^{(n,m)}.
\end{equation}}

This symmetric indicator assigns to both spikes a $+1$ in case the two spikes
are in the correct order, i.e. the spike from the spike train with the lower
spike train index is leading the coincidence, and a $-1$ in the opposite case.
Once more the value $0$ is obtained when there is no coincident spike in the other
spike train or when the two coincident spikes are absolutely identical.

The multivariate profile $E(t_k)$, again obtained similarly to
Eq.~\ref{eq:Multi-Profile}, is also normalized between $1$ and $-1$ and 
the extreme values are obtained for a coincident event covering all spike trains 
with all spikes emitted in the order from first (last) to last (first) spike 
train, respectively (see the first two and the last four events in Fig.
\ref{fig:SPIKE_Order_80_Standard_Example_1_Profiles}).
It can be $0$ either if a spike is not a part of any coincidences or if the
order is such that correctly and incorrectly ordered spike train pairs cancel
each other.
Again, $C_k$ is an upper bound for the absolute value of $E_k$.

\subsection{\label{ss:Synfire-Indicator} Synfire Indicator}

In contrast to the SPIKE-Order profile $D_k$, for the Spike Train Order profile
$E_k$ it does make sense to define an average value, which we term the Synfire
Indicator:
\begin{equation} \label{eq:Synfire Indicator-E}
	F = \frac{1}{M} \sum_{k=1}^M E(t_k).
\end{equation}

The interpretation is very intuitive.
The Synfire Indicator $F$ quantifies to what degree the spike trains in their
current order resemble a perfect synfire pattern.
It is normalized between $1$ and $-1$ and attains the value $1$ ($-1$) if the
spike trains in their current order form a perfect (inverse) synfire pattern.
This means that all spikes are coincident with spikes in all other spike trains and
\highlight{that all orders from leading (following) to following (leading) spike
consistently reflect the order of the spike trains.}

It is $0$ either if the spike trains do not contain any coincidences at all or
if among all spike trains there is a complete symmetry between leading and
following spikes. 

The Spike Train Order profile $E(t_k)$ for our example is shown in Fig. 
\ref{fig:SPIKE_Order_80_Standard_Example_1_Profiles}c.
In this case the order of spikes within an event clearly matters.
The Synfire Indicator $F$ is slightly negative indicating that the current
order of the spike trains is actually closer to an inverse synfire pattern.

%
%
\begin{figure}
    \includegraphics[width=85mm]{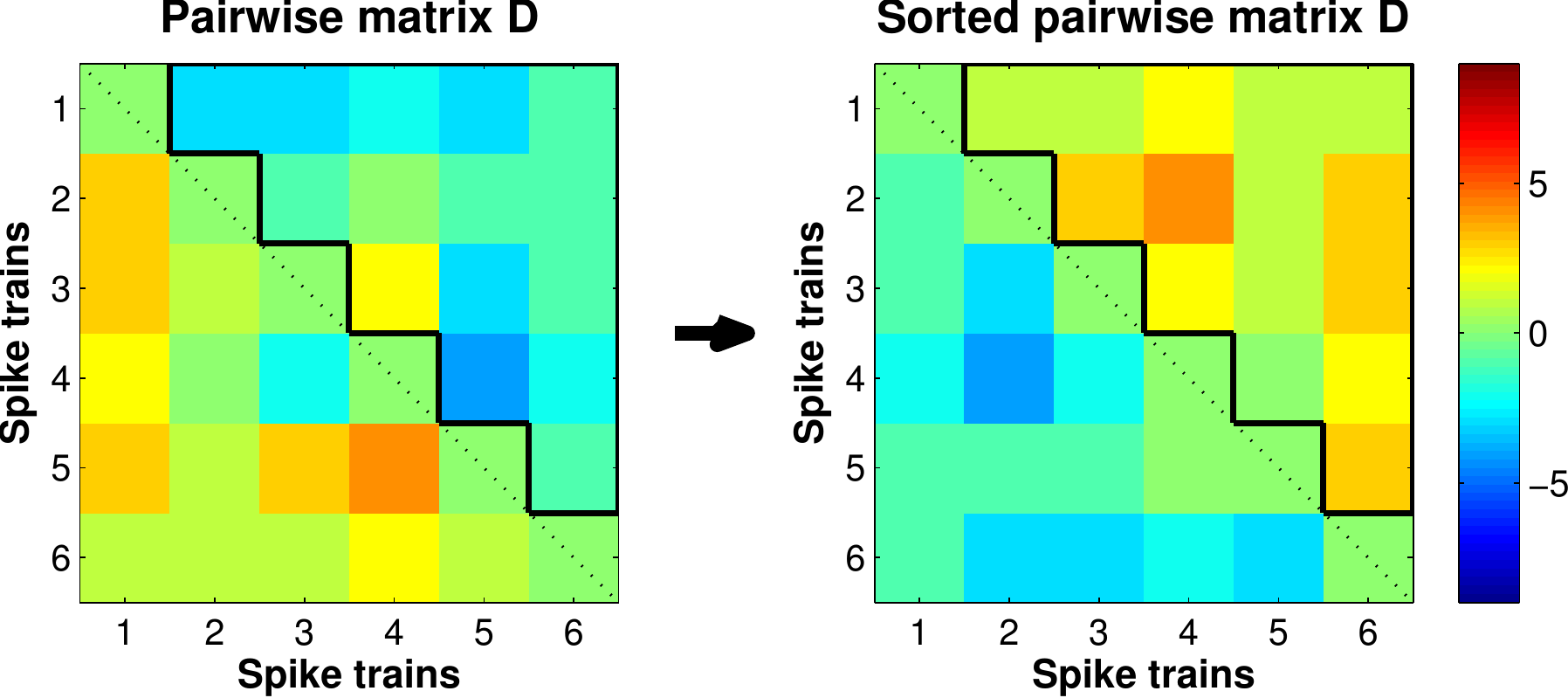}
    \caption{\abb\label{fig:SPIKE_Order_80_Standard_Example_2_Matrices}
Pairwise cumulative SPIKE-Order matrix $D$ before (left) and after (right)
sorting for the example dataset from Fig.
\ref{fig:SPIKE_Order_80_Standard_Example_1_Profiles}.
The upper triangular matrix $D^{(n<m)}$, marked in black, is used to calculate 
the Synfire Indicator $F$, for both the unsorted spike trains ($F_u$, left) and 
the sorted spike trains ($F_s$, right).
The thick black arrow in between the two matrices indicates the sorting process.
}
\end{figure}
%
%

Given a set of spike trains we now would like to sort the spike trains from
leader to follower such that the set comes as close as possible to a synfire
pattern.
To do so we have to maximize the overall number of correctly ordered
coincidences and this is equivalent to maximizing the Synfire Indicator $F$.
However, it would be very difficult to achieve this maximization by means of
the multivariate profile $E(t_k)$.
Clearly, it is more efficient to sort the spike trains based on a pairwise
analysis of the spike trains.
The most intuitive way is to use the anti-symmetric cumulative SPIKE-Order
matrix
\begin{equation}
    D^{(n,m)} = \sum_i D_i^{(n,m)}
\end{equation}
which sums up orders of coincidences from the respective pair of spike trains 
only and quantifies how much spike train $n$ is leading spike train $m$ (Fig. 
\ref{fig:SPIKE_Order_80_Standard_Example_2_Matrices}). 

Hence if $D^{(n,m)}>0$ spike train $n$ is leading $m$, while $D^{(n,m)}<0$ means 
$m$ is leading $n$.
If the current Spike Train Order is consistent with the synfire property, we 
thus expect that $D^{(n,m)} > 0$ for $n<m$ and $D^{(n,m)} < 0$ for $n>m$. 
Therefore, we construct the overall SPIKE-Order as
\highlight{\begin{equation} \label{eq:synfire_ind_D}
 	D_< = \sum_{n<m} D^{(n,m)},
\end{equation}}
i.e.\ the sum over the upper right tridiagonal part of the matrix $D^{(n,m)}$.

After normalizing by the overall number of possible coincidences, we arrive at
a second more practical definition of the Synfire Indicator:
\highlight{\begin{equation} \label{eq:Synfire Indicator-D}
	F = \frac{2 D_<}{(N-1) M}. 
\end{equation}}
The value is identical to the one of Eq. \ref{eq:Synfire Indicator-E}, only
the temporal and the spatial summation of coincidences (i.e., over the profile
and over spike train pairs) are performed in the opposite order.

Having such a quantification depending on the order of spike trains, we can 
introduce a new ordering in terms of the spike train index permutation $\varphi(n)$.
The overall Synfire Indicator for this permutation is then denoted as
$F_\varphi$.
Accordingly, for the initial (\textbf{u}nsorted) order of spike trains 
$\varphi_u$ the Synfire Indicator is denoted as $F_u = F_{\varphi_u}$.

The aim of the analysis is now to find the optimal (\textbf{s}orted) order
$\varphi_s$ as the one resulting in the maximal overall Synfire Indicator
$F_s = F_{\varphi_s}$:
\begin{equation}
	\varphi_s: F_{\varphi_s} = \max_\varphi \{F_\varphi\} = F_s.
\end{equation}
This Synfire Indicator for the sorted spike trains quantifies how close spike
trains can be sorted to resemble a synfire pattern, i.e., to what extent
coinciding spike pairs with correct order prevail over coinciding spike pairs
with incorrect order.
Unlike the Synfire Indicator for the unsorted spike trains $F_u$, the optimized
Synfire Indicator $F_s$ can only attain values between $0$ and $1$ (any order that
yields a negative result could simply be reversed in order to obtain the same
positive value).
For a perfect synfire pattern we obtain $F_s=1$, while sufficiently long
Poisson spike trains without any synfire structure yield $F_s\approx 0$.

The complexity of the problem to find the optimal Spike Train Order is 
similar to the well-known travelling salesman problem \citep{Applegate11}. 
For $N$ spike trains there are $N!$ permutations~$\varphi$, so for large numbers
of spike trains finding the optimal Spike Train Order $\varphi_s$ is a non-trivial
problem and brute-force methods such as calculating the $F_\varphi$-value for 
all possible permutations are not feasible. 
Instead, one has to make use of methods such as parallel tempering
\citep{Earl05} or simulated annealing \citep{Dowsland12} to search for the
optimal order.
Here we choose simulated annealing, a probabilistic technique which
approximates the global optimum of a given function in a large search space.
In our case this function is the Synfire Indicator $F_\varphi$ (which we would
like to maximize) and the search space is the permutation space of all spike
trains.
We start with the $F_u$-value from the unsorted permutation and then visit
nearby permutations using the fundamental move of exchanging two neighboring
spike trains within the current permutation.
\highlight{The update of the Synfire Indicator when exchanging the spike
trains $k$ and $k+1$ is simply given by $\Delta F = -2D^{(k, k+1)}$.}
All moves with positive $\Delta F$ are accepted while the likelihood of 
accepting moves with negative $\Delta F$ is decreased along the way according to 
a standard slow cooling scheme.
The procedure is repeated iteratively until the order of the spike trains no
longer changes or until a predefined end temperature is reached.

\begin{figure}
    \includegraphics[width=85mm]{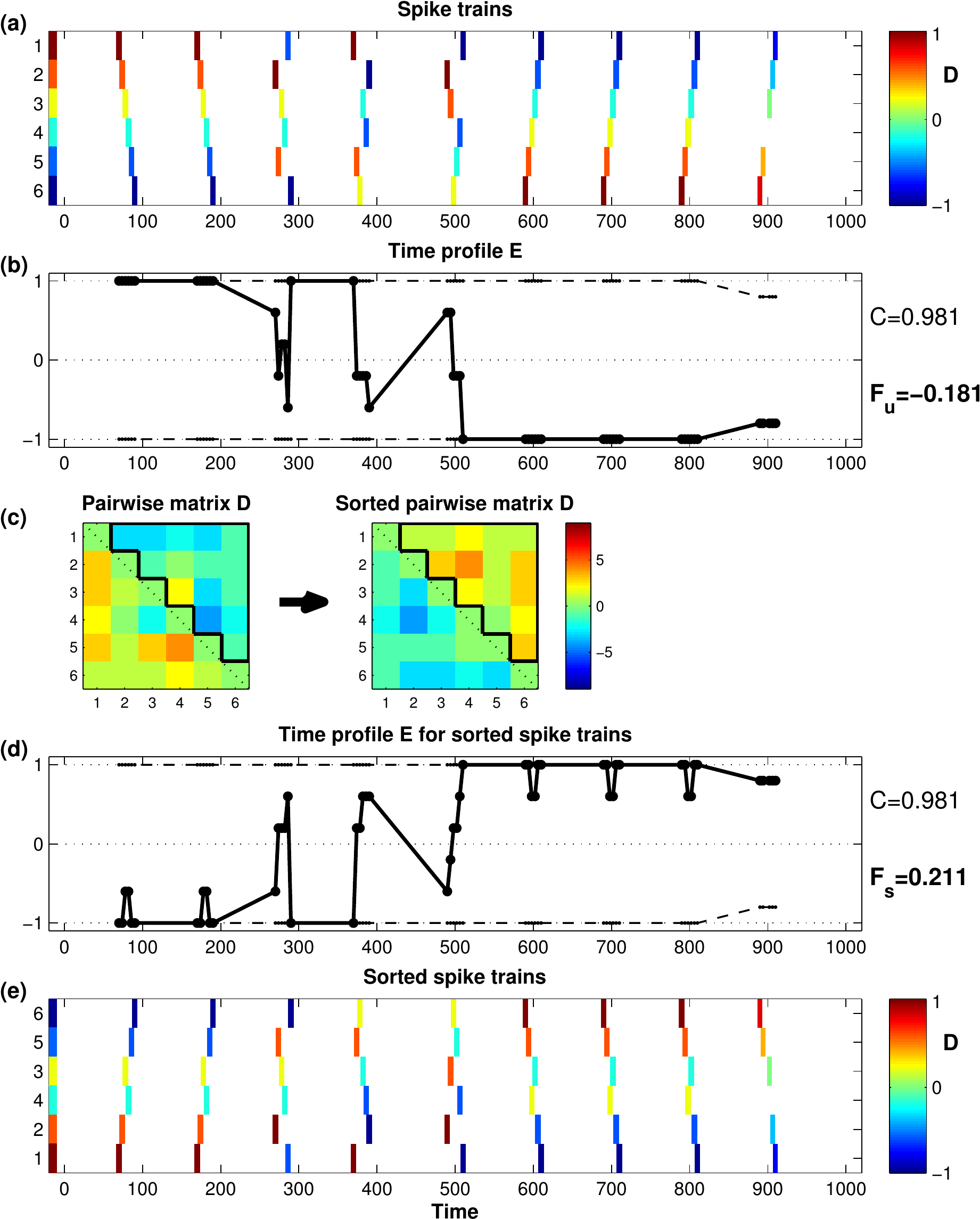}
    \caption{\abb\label{fig:SPIKE_Order_80_Standard_Example-3}
Complete illustration of SPIKE-Order using our example dataset from Fig.
\ref{fig:SPIKE_Order_80_Standard_Example_1_Profiles}.
(a) Unsorted spike trains with the spikes color-coded according to the value of
the SPIKE-Order $D(t_k)$.
(b) Spike Train Order profile $E(t_k)$. The Synfire Indicator $F_u$ for the
unsorted spike trains is slightly negative.
(c) Pairwise SPIKE-Order matrix $D$ before and after sorting. The optimal order
maximizes the upper triangular matrix.
(d) Spike Train Order profile $E(t_k)$ and its average values, the Synfire
Indicator $F_s$ for the sorted spike trains.
(e) Sorted spike trains.}
\end{figure}

In Fig. \ref{fig:SPIKE_Order_80_Standard_Example-3} we show the complete
SPIKE-order analysis including the results for the sorted spike trains.
The sorting of the spike trains maximizes the Synfire Indicator as reflected by
both the normalized sum of the upper right half of the pairwise cumulative
SPIKE-Order matrix (Eq.\ref{eq:Synfire Indicator-D}, Fig.
\ref{fig:SPIKE_Order_80_Standard_Example-3}c) and the average value of the
Spike Train Order profile $E (t_k)$ (Eq.\ref{eq:Synfire Indicator-E},
Fig. \ref{fig:SPIKE_Order_80_Standard_Example-3}d).
Finally, the sorted spike trains in Fig. \ref{fig:SPIKE_Order_80_Standard_Example-3}e
are now ordered such that the first spike trains have predominantly high values
(red) and the last spike trains predominantly low values (blue) of $D (t_k)$.

%
%
%
%

The complete analysis returns results consisting of several levels of
information.
Time-resolved (local) information is represented in the spike-coloring and in
the profiles $D$ and $E$.
The pairwise information in the SPIKE-order matrix reflects the leader-follower
relationship between two spike trains at a time.
The Synfire Indicator $F$ characterizes the closeness of the dataset as a whole
to a synfire pattern, both for the unsorted ($F_u$) and for the sorted ($F_s$)
spike trains.
Finally, the sorted order of the spike trains is a very important result in
itself since it identifies the leading and the following spike trains.

\subsection{\label{ss:Statistical-Significance} Statistical significance}

As a last step in the analysis we evaluate the statistical significance of the 
optimized Synfire Indicator $F_s$.
\highlight{What we would like to estimate is the likelihood that for the given total number
of coincidences the prevalence of correctly ordered spike pairs (as quantified
by the optimized Synfire Indicator) could have been obtained by chance.
If all coincident spike pairs would be independent, the probability distribution
would be strictly binomial and we could calculate this likelihood analytically.
However, the pairwise spike orders in coincident events involving multiple spike trains
are not independent from each other, and so instead we estimate the likelihood
numerically using a set of carefully constructed spike order surrogates.}

\begin{figure}
\includegraphics[width=85mm]{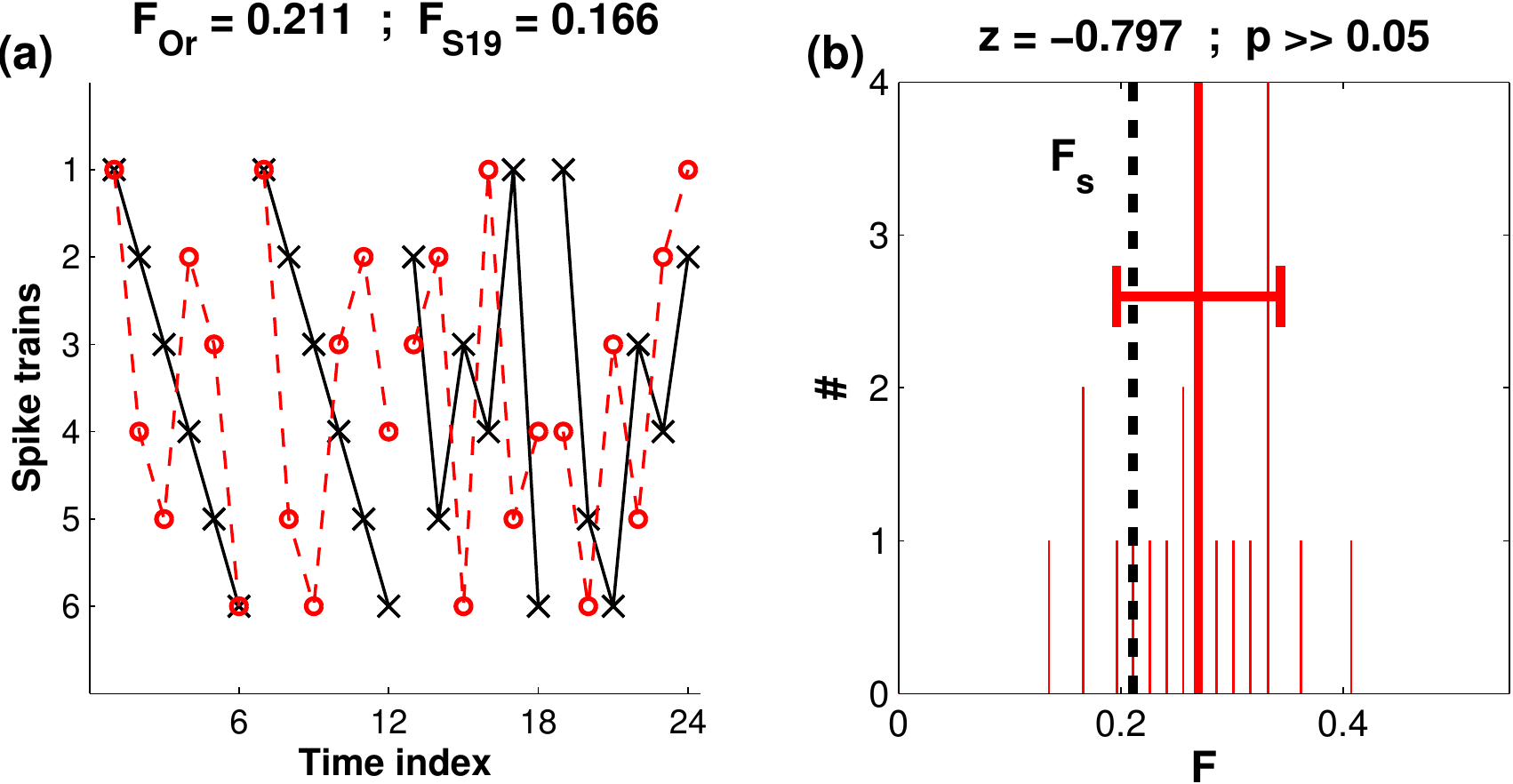}
    \caption{\abb\label{fig:SPIKE_Order_Surro_80_Standard_Example_4_Surro-Histo} 
Statistical significance:
Surrogate analysis for the example dataset from Fig. 
\ref{fig:SPIKE_Order_80_Standard_Example_1_Profiles}.
(a) Spike order patterns for original (black) and one randomized surrogate
(red).
For clarity only the first four events are shown.
For the first two events the synfire-order of the original is destroyed in the
surrogates whereas for the next two events both sequences are equally
unordered.
(b) Histogram for $19$ surrogates.
Thick lines denote mean and standard deviation.
Since the value for the original dataset (black) is not maximum, the optimally
sorted spike trains do not exhibit a statistically significant synfire pattern.
}
\end{figure}

\highlight{For each surrogate (Fig. \ref{fig:SPIKE_Order_Surro_80_Standard_Example_4_Surro-Histo}a)
we maintain the coincidence structure of the original spike trains by preserving the
SPIKE-Synchronization values of every individual spike.
However, we destroy the spike order patterns by swapping the order of the two spikes in a
sufficient number of randomly selected coincident spike pairs.
Note that the generation of surrogates takes place not on the level of spike times
but on the level of order values (the x-axis in
Fig. \ref{fig:SPIKE_Order_Surro_80_Standard_Example_4_Surro-Histo}a is labeled 'time index',
not 'time').
Spike trains with swapped spike times would have different interspike intervals, and this
would alter the results of the coincidence criterion in Eq. \ref{eq:Coincidence-MaxDist} 
and change the value of SPIKE-Synchronization.
This in turn would make the desired evaluation of pure spike order effects difficult.}
%
%
%
%

\highlight{In the implementation, from one spike order surrogate to the next the number of
spike order swaps is set to the number of coincident spikes in the spike train set,
such that all possible spike order patterns can be reached.
Only for the first surrogate, since it starts from the original spike trains,
we swap twice as many coincidences in order to account for transients.}
After each swap we take extra care that all other spike orders that are affected by the 
swap are updated as well.
For example, if a swap changes the order between the first and the third spike
in an ordered sequence of three spikes, we also swap both the order between the
first and the second as well as the order between the second and the third spike.

For each spike train surrogate we repeat exactly the same optimization procedure 
in the spike train permutation space that is done for the original dataset.
The original Synfire Indicator is deemed significant if it is higher than the 
Synfire Indicator obtained for all of the surrogate datasets (this case will be 
marked by two asterisks).
Here we use $s = 19$ surrogates for a significance level of
$p^* = 1/(s+1) = 0.05$.
As a second indicator we state the z-score, e.g., the deviation of the original
value $x$ from the mean $\mu$ of the surrogates in units of their standard
deviation $\sigma$:
\begin{equation}
    z = \frac{x-\mu}{\sigma}.
\end{equation}

Results of the significance analysis for our standard example are shown in the 
histogram in Fig. \ref{fig:SPIKE_Order_Surro_80_Standard_Example_4_Surro-Histo}b.
In this case the absolute value of the z-score is smaller than one and the
p-value is larger than $p^*$ and the result is thus judged as statistically
non-significant.

In case the initial sorting of the spike trains is used to test a specific 
hypothesis there also exists a straightforward procedure to test the statistical 
significance of the Synfire Indicator $F_u$ for the unsorted spike trains.
In this case no optimization of the Synfire Indicator is required, rather the 
Synfire Indicator $F_u$ for the initial sorting is compared against Synfire 
Indicators obtained for random permutations of the spike trains.
This kind of significance test will be used in Section \ref{ss:Results-Neuro}.


%
%

\section{\label{s:Results} Results}

In the following we apply our new algorithm to artificially generated datasets 
(Section \ref{ss:Results-Sim}), neurophysiological data (Section 
\ref{ss:Results-Neuro}) and, finally, climate data from the El Ni\~{n}o 
phenomenon (Section \ref{ss:Results-Climate}). For the Figures we use the same
full layout introduced in Fig. \ref{fig:SPIKE_Order_80_Standard_Example-3} to
which we add the significance analysis of Fig. 
\ref{fig:SPIKE_Order_Surro_80_Standard_Example_4_Surro-Histo}b.

\subsection{\label{ss:Results-Sim} Application to artificially generated data}

%
%
\begin{figure}
    \includegraphics[width=85mm]{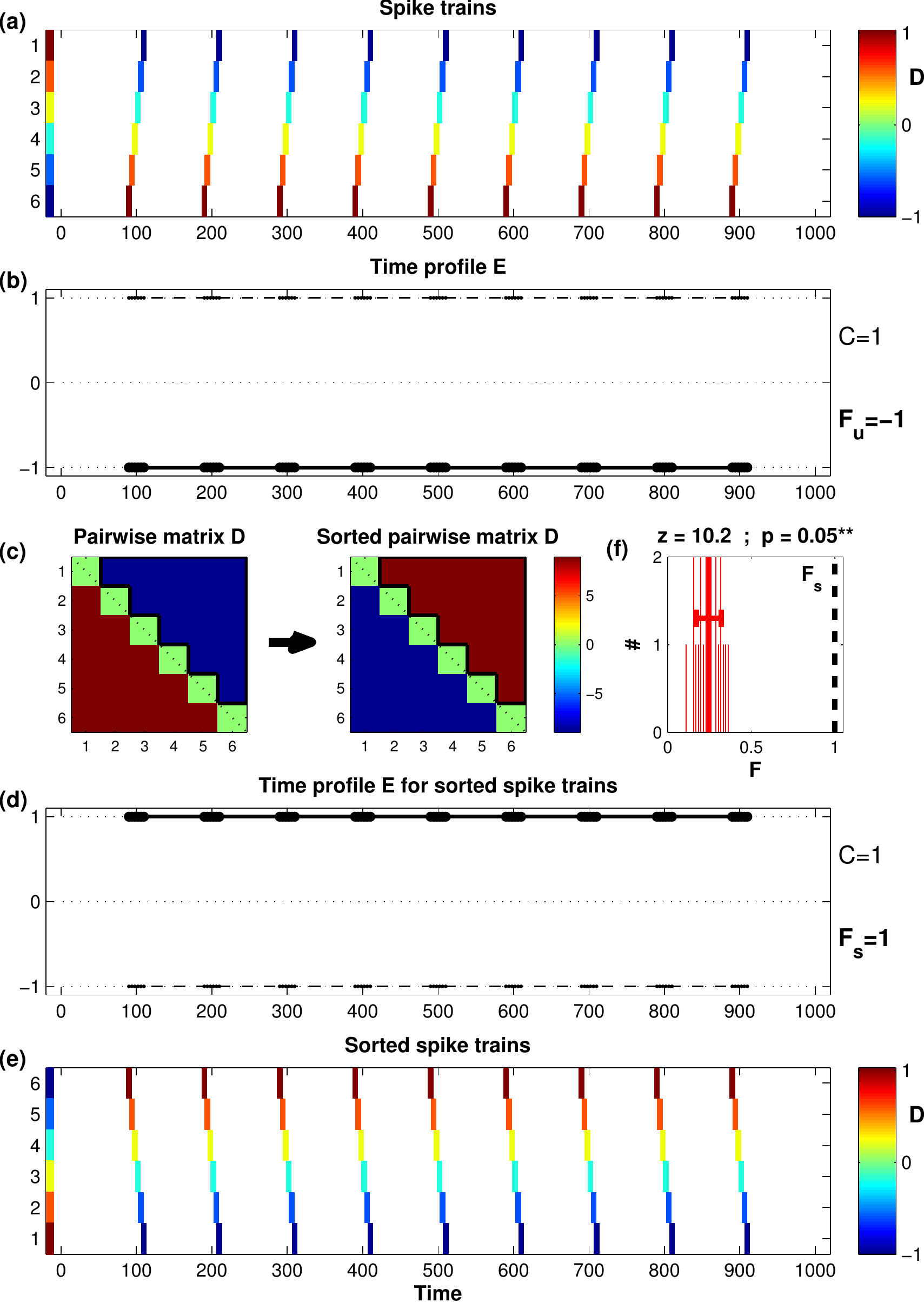}
    \caption{\label{fig:SPIKE_Order_Surro_2_SynFire_Inv} SPIKE- and Spike Train 
Order analysis for a perfect inverse synfire pattern. The plot follows the 
layout of Fig. \ref{fig:SPIKE_Order_80_Standard_Example-3} with the histogram of 
the surrogate test (see Fig. 
\ref{fig:SPIKE_Order_Surro_80_Standard_Example_4_Surro-Histo}b) for statistical 
significance added as subplot f. For the unsorted spike trains a minimal Synfire 
Indicator of $F_u=-1$ is obtained, while sorting results in the maximum value of 
$F_s=1$. According to the surrogate test the statistical significance of the 
result is very high.}
\end{figure}
%
%

We start with examples covering the two extreme cases of a perfect synfire 
pattern and a completely random spike train set.
First, in Fig. \ref{fig:SPIKE_Order_Surro_2_SynFire_Inv} we apply the algorithm
to a perfect inverse synfire pattern for which the spike trains are initially
sorted from follower to leader.
Therefore, the Synfire Indicator of the unsorted spike trains yields its
minimum value of $F_u = -1$.
Sorting just reverses the order of the spike trains and in consequence the
maximum value of $F_s = 1$ is obtained.
Any shuffling of spike orders necessarily destroys the synfire pattern and thus
leads to much lower values of the Synfire Indicator.
Accordingly, the surrogate test (Fig. \ref{fig:SPIKE_Order_Surro_2_SynFire_Inv}f)
shows that the statistical significance of the original Synfire Indicator is
very high.

%
%
\begin{figure}
    \includegraphics[width=85mm]{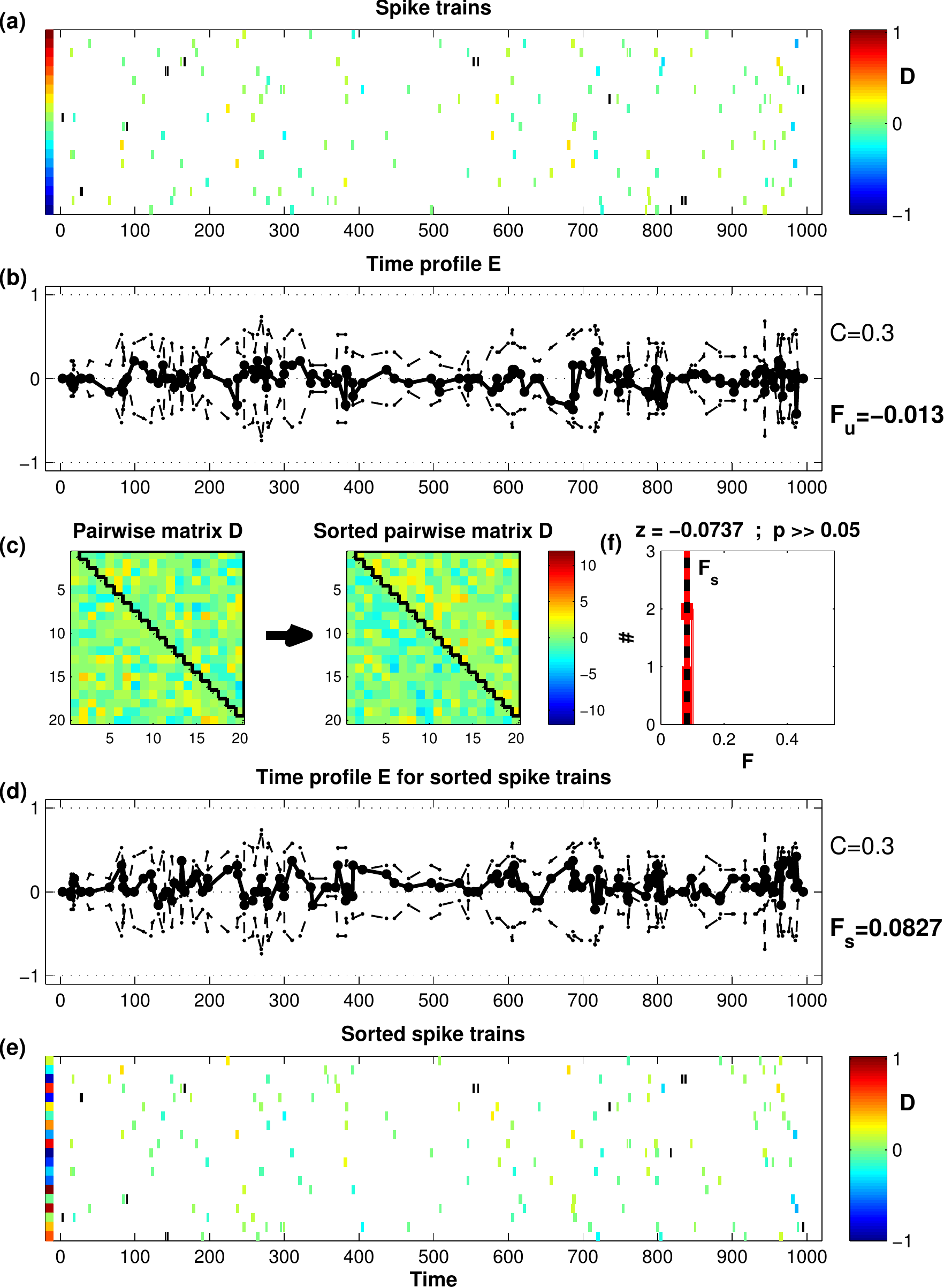}
    \caption{\label{fig:SPIKE_Order_Surro_51_Poisson} SPIKE- and Spike Train 
Order analysis for $20$ Poisson spike trains.
Since the number of spike trains is too large to label the spike trains in
the top and in the bottom subplot with numbers we use color coding at the left
side to label them.
Both before and after sorting the Synfire Indicator is very close to zero.
The surrogate analysis reveals the result to be non-significant.}
\end{figure}
%
%

The other extreme case is Poisson spike trains (Fig. 
\ref{fig:SPIKE_Order_Surro_51_Poisson}) for which the arrival times of spikes
are completely random and without any preferred order.
For this realization the Synfire Indicator $F_u$ for the unsorted spike trains
happens to be slightly negative indicating that the spike trains are closer to
an inverse synfire pattern than to a synfire pattern.
The absolute value $F_s$ after sorting is higher.
The fact that both of these values are non-zero is due to the finite size
effect caused by the limited number of spikes.
For more and more spike trains and/or spikes the expectation value even for the
sorted case would converge towards zero.
As expected, the surrogate test shows that the order for the original 
spike trains is not statistically distinct from the order of the surrogate
spike trains (there is no preferred order that can be destroyed by the
shuffling) and, accordingly, the value of the original Synfire Indicator is
revealed to be clearly non-significant.
%
%
\begin{figure}
    \includegraphics[width=85mm]{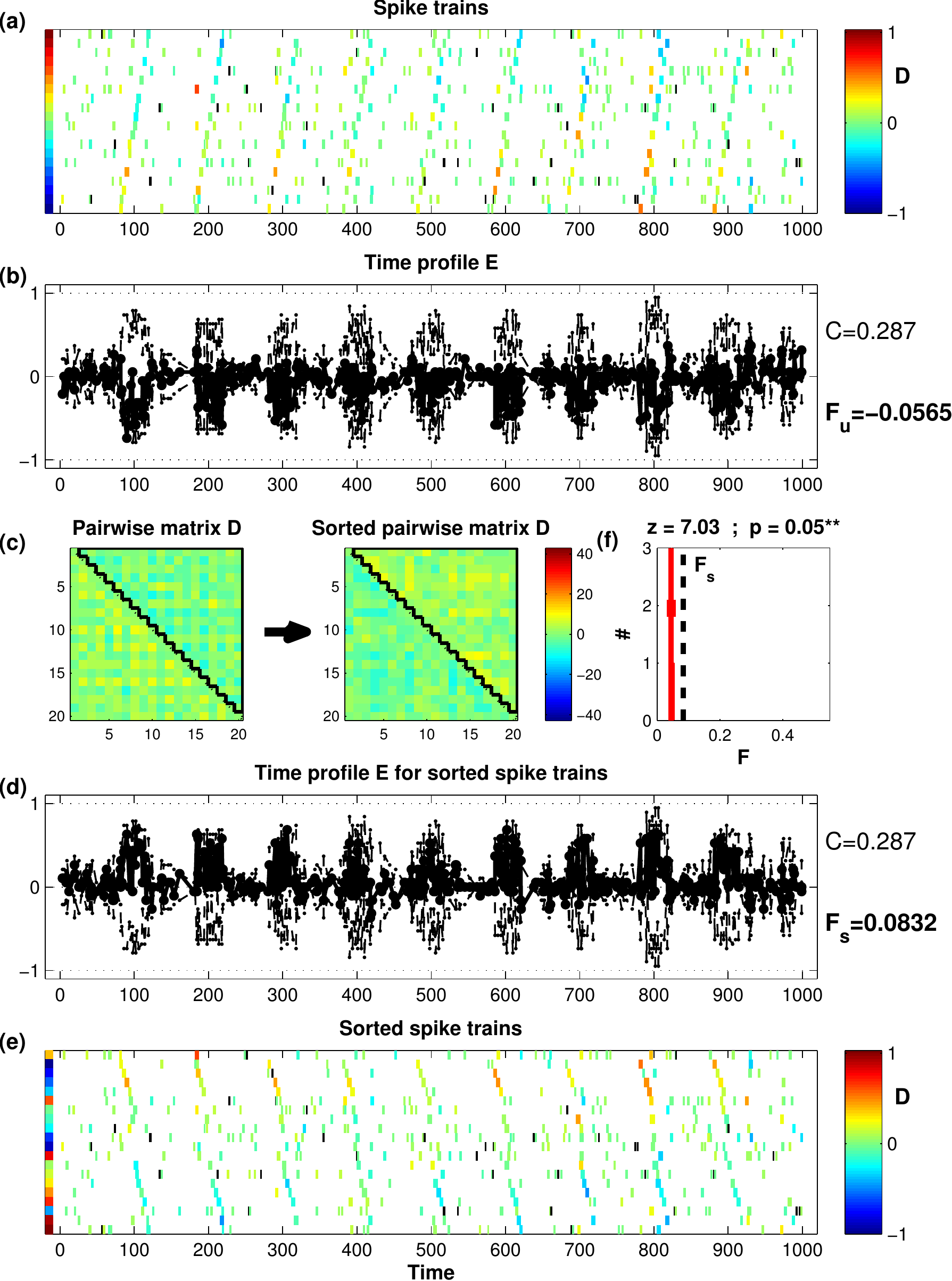}
    \caption{\label{fig:SPIKE_Order_Surro_57_Mix_Poisson_SF}
SPIKE- and Spike Train Order analysis for Poisson spike train interspersed with
\highlight{spike trains that contain random spikes but also} a perfect inverse
synfire pattern.
The order contained within the synfire pattern spike train is distinct enough
to make the Synfire Indicator for the sorted spike trains statistically
significant.}
\end{figure}
%
%

The third example in Fig. \ref{fig:SPIKE_Order_Surro_57_Mix_Poisson_SF} shows
a mixture of these two extremes, Poisson spike train interspersed with spike 
trains \highlight{that contain a perfect inverse synfire pattern (plus random
spikes)}.
Sorting the spike trains restores the correct order of the synfire pattern
spike trains within the Poisson spike train.
The Synfire Indicator for the sorted spike trains $F_s$ for this mixed example
is actually almost identical to the value obtained for the Poisson spike trains in
Fig. \ref{fig:SPIKE_Order_Surro_51_Poisson}, but this time the surrogate test
reveals the value to be highly significant.
These two examples combined illustrate nicely that the Synfire Indicator and the
surrogate analysis provide complementary information.
In the mixture example of Fig. \ref{fig:SPIKE_Order_Surro_57_Mix_Poisson_SF}
there are many more random Poisson spikes than ordered synfire pattern spikes.
\highlight{According to the Synfire Indicator, these two types of spikes together
appear to be as ordered as the spikes of the shorter but purely random Poisson
spike trains in Fig. \ref{fig:SPIKE_Order_Surro_51_Poisson}.}
However, the Synfire Indicator is strongly influenced by the statistics of the dataset
and thus is in itself not sufficient to reliably compare two datasets with
widely different number of spike trains and spikes.
The surrogate analysis, on the other hand, can be used to compare datasets of
different size since by preserving the spike numbers in the surrogates it
explicitly takes the statistics of each dataset into account.

\subsection{\label{ss:Results-Neuro} Application to neuroscience}

In order to apply the Spike Train Order algorithm to real neurophysiological 
data, we analyzed data recorded via fast multicellular calcium imaging in acute 
CA3 hippocampal brain slices from juvenile mice.
In the juvenile hippocampus, the CA3 region is the origin of a stereotypical
network phenomenon of wavelike propagating activity termed giant depolarizing
potentials (GDPs \cite{BenAri89}).
In previous studies, GDPs have been used to investigate the topology of
networks and the role of hub cells \citep{Bonifazi09} as well as to reveal the
deterministic and stochastic processes underlying spontaneous, synchronous
network bursts \citep{Takano12}.
Due to the distinct architecture and the repetitive nature of the GDPs this
experimental setup offers a very suitable test case for our synfire pattern
analysis \highlight{(for more background and a detailed description of the experimental
methods refer to Appendix \ref{App-s:HM-Data})}. 

%
%
\begin{figure}
    \includegraphics[width=85mm]{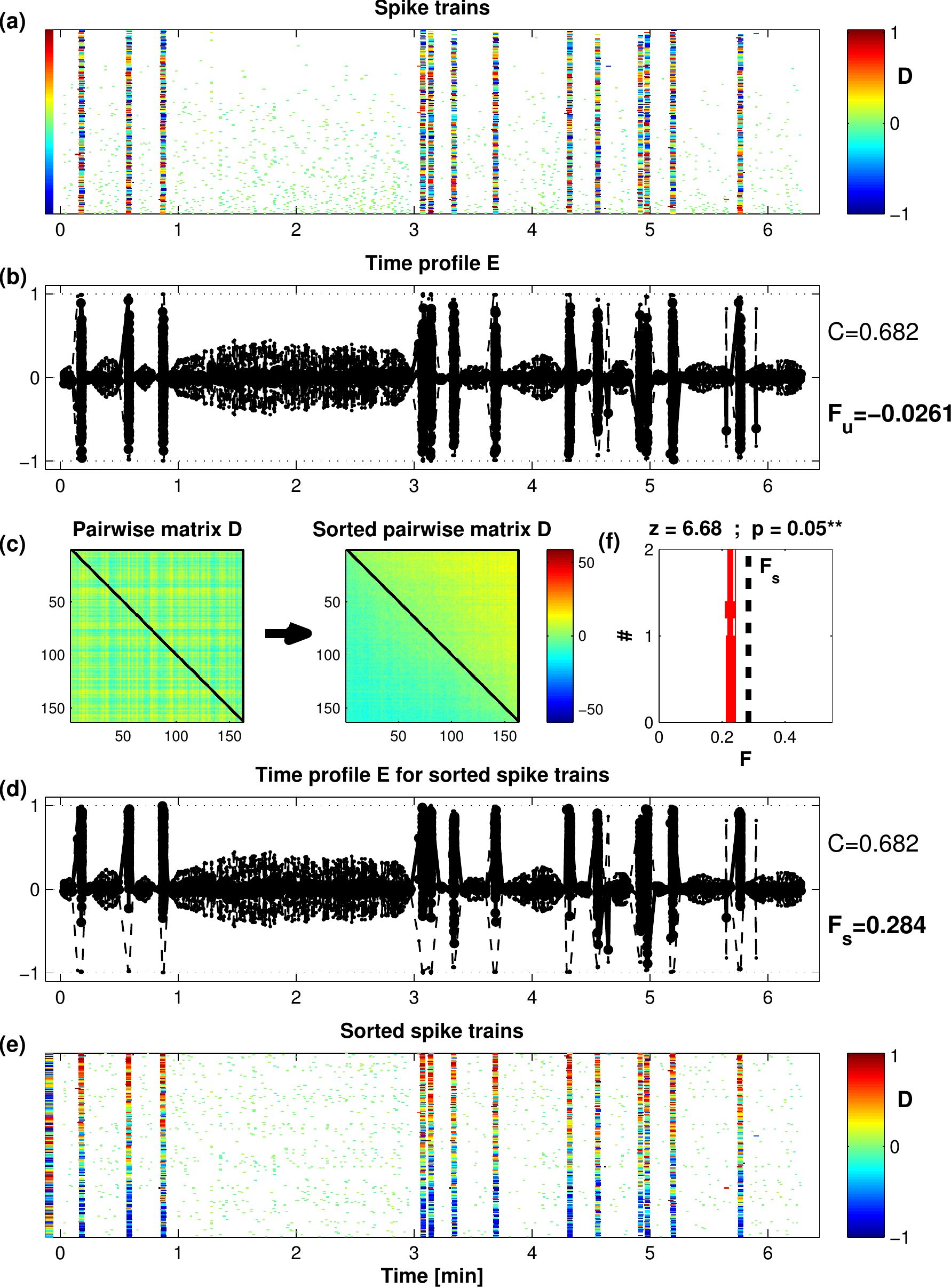}
    \caption{\label{fig:SPIKE_Order_Surro_109_HM_5_1_SN}
SPIKE-Order for real data recorded in an acute hippocampal slice from a
juvenile mouse. \highlight{Note how the color-coding of the spikes according to their
SPIKE-order $D$ helps to overcome the low temporal resolution of the Figure
and to resolve the spike order within the GDPs.}
(a) Initially the spike trains are sorted according to their firing rate
starting with the most sparse spike trains. \highlight{The messy color-patterns
reveal that this is completely uncorrelated to the spike order within the GDPs.} 
(f) After sorting, there is a fairly consistent transition from spike trains with
predominantly leading spikes (red) in the GDPs to spike trains with
predominantly following spikes (blue).
}
\end{figure}
%
%

The first dataset analyzed in Fig. \ref{fig:SPIKE_Order_Surro_109_HM_5_1_SN} 
includes $13$ GDPs over a bit more than $6$ minutes.
Almost all GDPs involve the whole network.
Here as for all other neurophysiological datasets analyzed initially the spike
trains are sorted according to their firing rate such that the sparsely spiking
neurons are on top and the most active neurons at the bottom
(Fig. \ref{fig:SPIKE_Order_Surro_109_HM_5_1_SN}a).
This specific sorting allows us to test the hypothesis that the neurons which
fire almost exclusively within the GDPs and are very sparse on background
activity might have a stronger role in initiating GDPs and tend to lead,
whereas the more regularly spiking neurons might tend to follow.
If this would be the case one would expect a very high value for the initial
Synfire Indicator $F_u$.
However, according to Fig. \ref{fig:SPIKE_Order_Surro_109_HM_5_1_SN}b the
actual value is very close to zero and actually slightly negative.
A statistical significance test using random permutations of spike trains (see
Section \ref{ss:Statistical-Significance}) indeed proves the Synfire Indicator
of the unsorted spike trains $F_u$ to be non-significant (result not shown).
A further indicator for this is the fact that the order of the sorted spike 
trains is very different from the initial order, as can be seen by comparing 
the color bars on the left of Fig. \ref{fig:SPIKE_Order_Surro_109_HM_5_1_SN}a 
and Fig. \ref{fig:SPIKE_Order_Surro_109_HM_5_1_SN}e.
The color-coding of the GDPs exhibits typically a slightly noisy transition
from leader (red) to follower (blue).
The Synfire Indicator for the sorted spike trains $F_s$ is also much higher
(Fig. \ref{fig:SPIKE_Order_Surro_109_HM_5_1_SN}d).
Finally, the surrogate analysis  (Fig.
\ref{fig:SPIKE_Order_Surro_109_HM_5_1_SN}f) shows this result to be highly 
significant.

%
%
\begin{figure}
    \includegraphics[width=85mm]{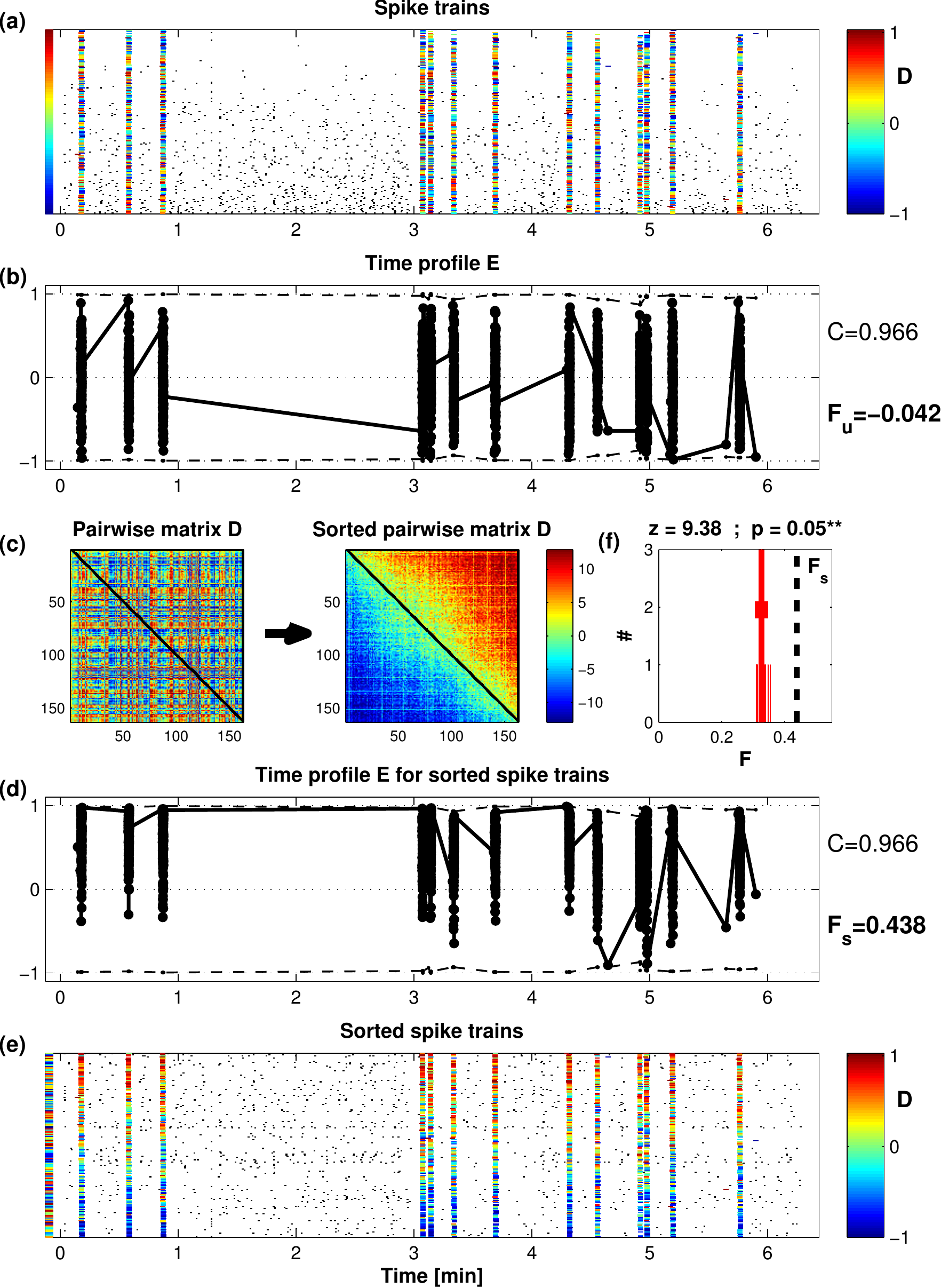}
    \caption{\label{fig:SPIKE_Order_Surro_159_HM_5_1_SN}
SPIKE-Order for the real data already analyzed in Fig.
\ref{fig:SPIKE_Order_Surro_109_HM_5_1_SN} but this time the analysis of
SPIKE-Order was restricted to spikes with a SPIKE-Synchronization value of at
least $0.7$.
This simple thresholding allows to focus the analysis on the reliable events
and to disregard all spikes between the events \highlight{(these are not colored and thus
remain black)}.
This results in an increase of the overall value of SPIKE-Order from $0.284$ to
$0.438$.
}
\end{figure}
%
%

However, the spiking in Fig. \ref{fig:SPIKE_Order_Surro_109_HM_5_1_SN} consists 
not only of the GDPs.
Most neurons exhibit at least to some extent spontaneous background activity,
the ones at the top of the initial sorting less than the ones at the bottom.
The spikes in this background activity are typically coincident with only few
other spikes and do not take part in any propagation patterns (note their green
color which indicates SPIKE-order values close to zero).
So in the context of our synfire pattern analysis this is just noise that leads
to a decrease of the Synfire Indicator.
There is a straightforward way to disregard these background spikes by setting a
threshold value $C_{thr}$ for the SPIKE-Synchronization profile $C(t_k)$.
Only spikes with a coincidence value higher than $C_{thr}$ are taken into
account, all other spikes are simply ignored.
The result of this background correction can be seen in Fig. 
\ref{fig:SPIKE_Order_Surro_159_HM_5_1_SN} for the same dataset already used in
Fig. \ref{fig:SPIKE_Order_Surro_109_HM_5_1_SN}.
Focusing the analysis on the reliable GDPs leads to an increase of the Synfire
Indicator from $0.284$ to $0.438$. 

%
%
\begin{figure}
    \includegraphics[width=85mm]{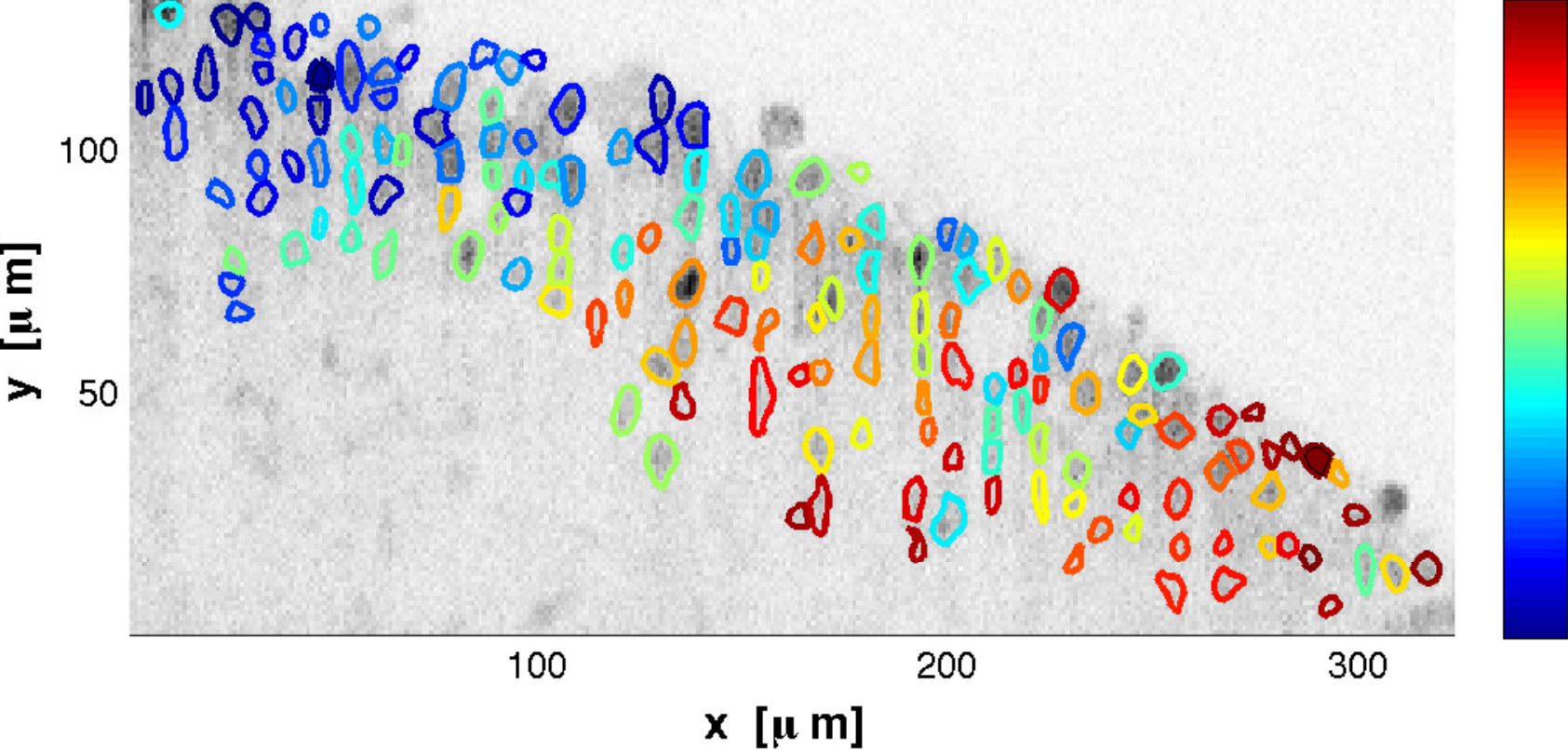}   
    \caption{\label{fig:2D-plot-ColorCoding-159}
Projection of the optimized Spike Train Order on the 2D-\highlight{photo} of the hippocampal
slice.
The Regions of Interest (ROIs) which denote filled and identified cells in the
CA3 region are color-coded from leader (index $1$, red) to follower
(index $163$, blue) using the optimized Spike Train Order of Fig. 
\ref{fig:SPIKE_Order_Surro_159_HM_5_1_SN}.
The very first leader (lower right) and the very last follower (upper left) are
marked by filled contours.
}
\end{figure}
%
%

As already mentioned before, one of the main results of our analysis is the 
sorted order of the spike trains itself.
For these neurophysiological data it allows to identify the leading and the
following neurons in the network and to project this information back on the
recording setup.
This is shown in Fig. \ref{fig:2D-plot-ColorCoding-159} where we have
color-coded the optimized Spike Train Order obtained in Fig.
\ref{fig:SPIKE_Order_Surro_159_HM_5_1_SN} within a 2D-plot of the neurons
recorded from the hippocampal slice.
For this example there appears to be a clear overall propagation from right to
left but there is also a considerable degree of \highlight{variability} which might be due to
a non-trivial connectivity within the network.


%
%
\begin{figure}
    \includegraphics[width=85mm]{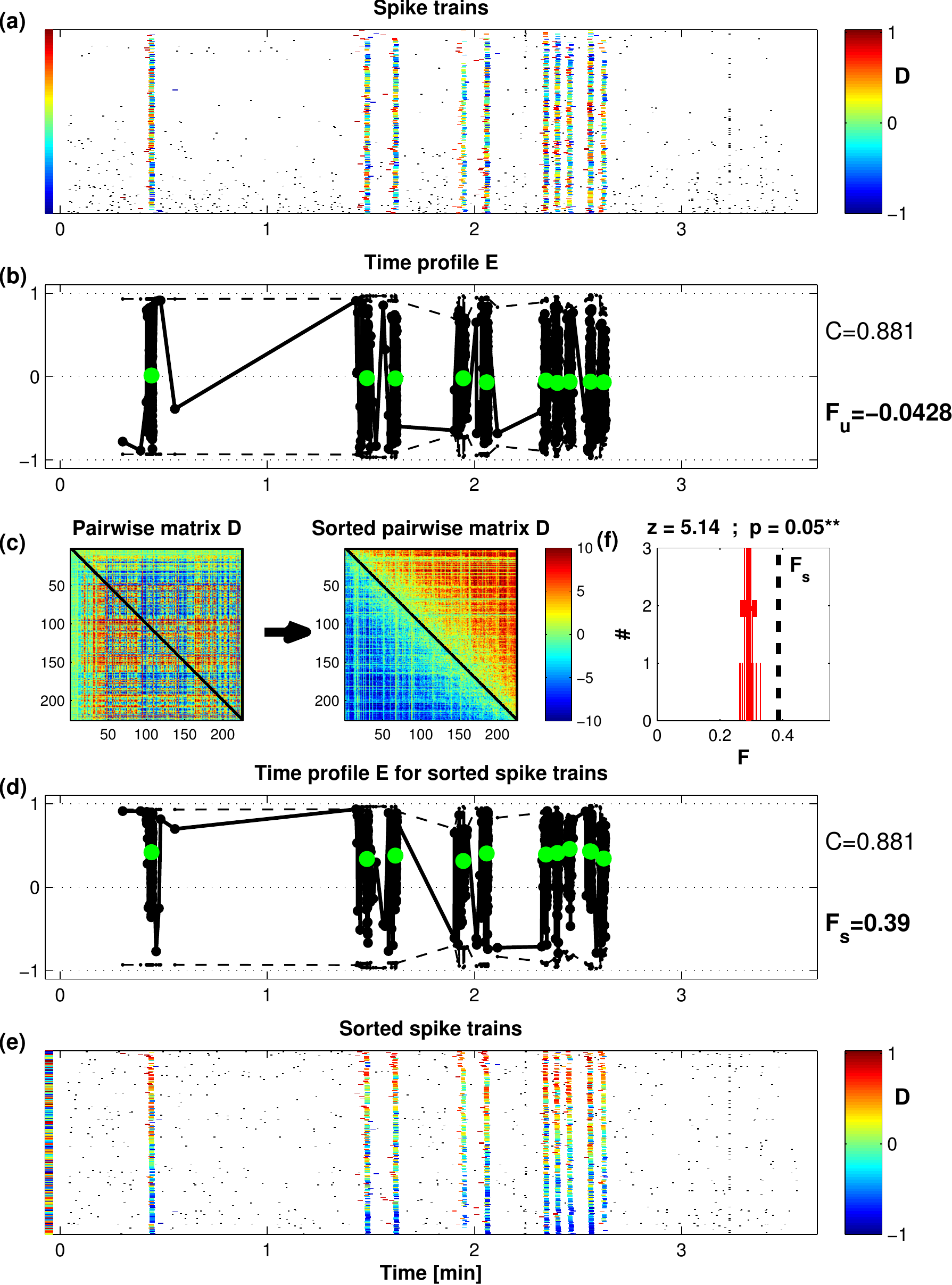}
    \caption{\label{fig:SPIKE_Order_Surro_175_HM_23_1_SN_Div}
SPIKE-Order for a second dataset for which again the analysis of SPIKE-Order
was restricted to spikes with a SPIKE-Synchronization value of at least $0.7$.
In this case the focus on the GDPs let the Synfire Indicator increase from
$0.296$ (result not shown) to $0.389$.
In addition, here we also calculated one average Spike Train Order value per GDP
(green points) which illustrates once more the time-resolved nature of the
method.
}
\end{figure}
%
%

In Fig. \ref{fig:SPIKE_Order_Surro_175_HM_23_1_SN_Div} we apply the SPIKE-order 
analysis to a second dataset recorded from a different slice, again focusing on 
the order within the global events only.
Here we also added one new feature, the mean value of the Spike Train Order
$E(t_k)$ for each global event (we use the maxima and minima of the SPIKE order
profile $D(t_k)$ to delineate the GDPs).
This again emphasizes the time-resolved nature of the SPIKE order and the Spike
Train Order indicators.

Overall, we have analyzed neurophysiological datasets from four hippocampal
slices exhibiting an average of $7.75$ GDPs.
We obtained an average value for SPIKE-Synchronization of $0.59$ before
focusing on the GDPs (as in Fig. \ref{fig:SPIKE_Order_Surro_109_HM_5_1_SN}) and
$0.92$ after (as in Fig. \ref{fig:SPIKE_Order_Surro_159_HM_5_1_SN}).
With or without this focus the Synfire Indicator for the initial spike train
sorting $F_u$ was very close to zero and in all cases proved to be
non-significant when tested against Synfire Indicators obtained for random
permutations of the Spike Train Order.
Since the initial sorting was based on overall firing rate of the neurons, this
signifies that the hypothesis that the low-firing neurons which are basically
only active during the GDPs might have a stronger role in initiating GDPs can
be rejected.
For the sorted spike trains the Synfire Indicator $F_s$ was $0.20$ for all
spikes and $0.42$ for the spikes within the GDPs only.
Suppressing the effect of the noisy background spikes in the analysis thus
leads to an average increase of the Synfire Indicator by about $110 \%$.
Finally, according to the surrogate analysis described in Section
\ref{ss:Statistical-Significance} the Synfire Indicator for the sorted 
spike trains $F_s$ yielded a statistically significant result for all datasets
analyzed.

So overall we can conclude that the GDPs recorded in brain slices from
juvenile mice are distinguished by a very high consistency of their
spatio-temporal propagation patterns.
\highlight{However, it is interesting to note that this consistency does not hold when
comparing different slices. In the datasets analyzed in this paper we find
examples of both propagation in the direction of CA2 as well as propagation towards
the dentate gyrus.
This is consistent with results reported in \citet{Takano12}.}

\subsection{\label{ss:Results-Climate} Application to climate data}

Although being developed mainly for neuroscientific data (spike train 
recordings), the Spike Train Order approach presented in this paper can be 
applied in many other contexts as well.
One particular field where event-based analysis is employed very frequently is 
climate science, see e.g.~\cite{Malik12, Boers14}.

\begin{figure}[t]
 \includegraphics[width=0.49\textwidth]{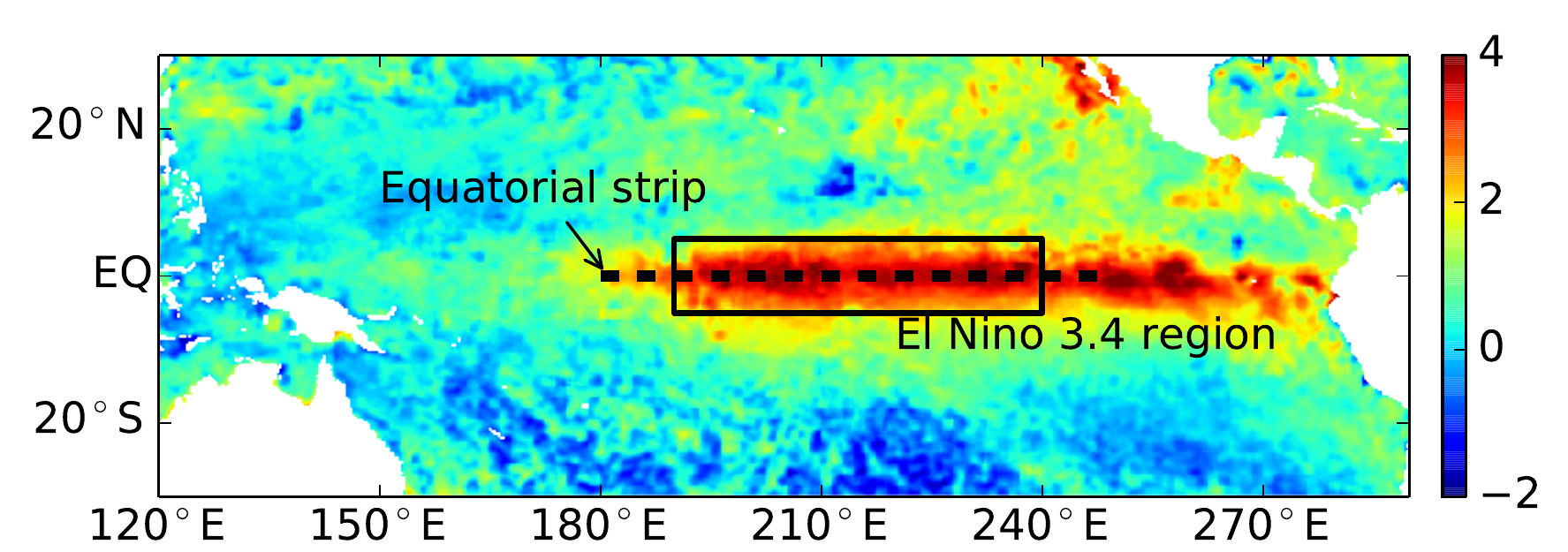}
 \caption{\elnino\ 3.4 region (rectangle) and the small strip around the equator 
used for the analysis in this work (dashed line). Color coding represents daily 
SST anomaly (in $\degC{}$) on November~11, 2015.}
 \label{fig:el_nino_region}
\end{figure}

\begin{figure}[t]
 \includegraphics[width=0.49\textwidth]{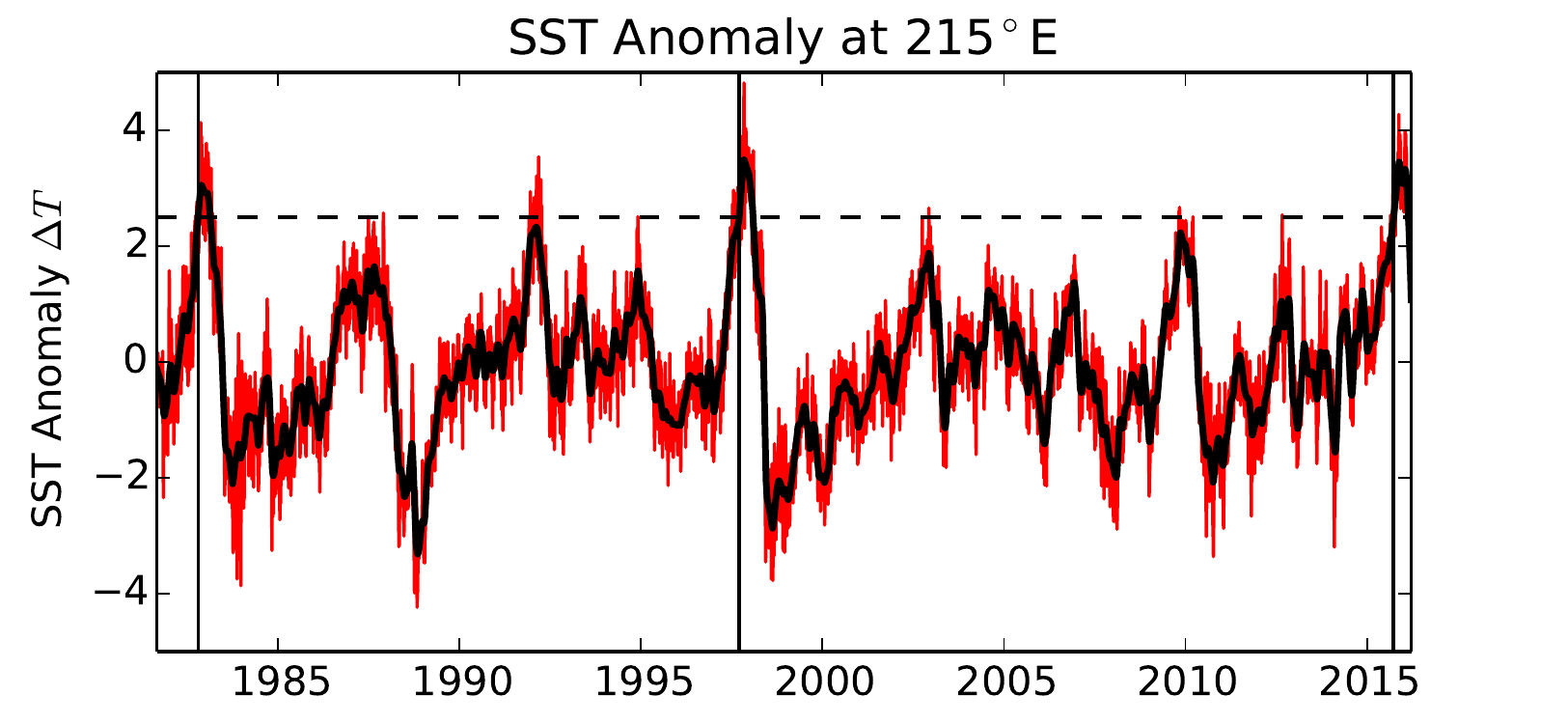}
 \caption{Exemplary time series of the SST anomaly at $\degE{215}$ before (red) 
and after (black) Gaussian smoothing. The horizontal dashed line corresponds to 
a temperature threshold of $\threshold=\degC{2.5}$ used for event detection and 
the vertical lines are the resulting events identified from upward threshold 
crossings of the smoothed time series.}
 \label{fig:time_series}
\end{figure}

In the following we will use the new Spike Train Order method to analyze the sea 
surface temperature (SST) in the central and eastern tropical Pacific Ocean to 
identify the propagation patterns connected with the warm phase of the \elnino\ 
Southern Oscillation (ENSO).
Predicting the occurrence and strength of \elnino s is very important due to 
the vast ecological and economical effects~\citep{McPhaden06} and therefore 
this phenomenon has been studied extensively in terms of both intensity  
(e.g. \cite{Rasmusson82, Yeh09}) and frequency (e.g. \cite{Qian11, An00}).
However, here we will not discuss \elnino\ prediction, but focus on the 
longitudinal propagation pattern of the \elnino\ events within the Pacific 
Ocean~\citep{Santoso13, Antico16}.
The region used here to analyze the SST is depicted in \figref{el_nino_region} 
(dashed line), and an examplary smoothed time profile for the center of the 
analyzed region ($\degE{215}$) is shown in \figref{time_series}.
Further details on the region, the dataset and data preprocessing (smoothing) is 
outlined in Appendix~\ref{App-s:ElNino-Data}.

\begin{figure}
\centering
 \includegraphics[width=0.4\textwidth]{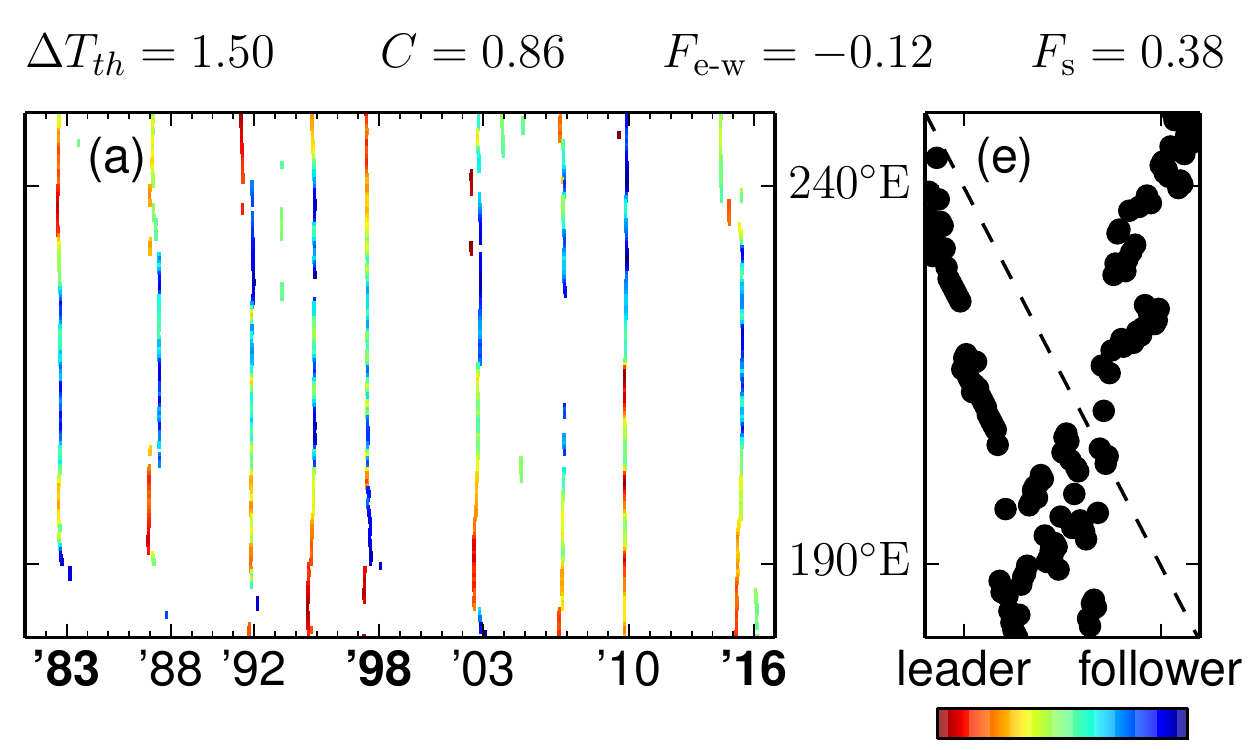}\\[.5em]
 \includegraphics[width=0.4\textwidth]{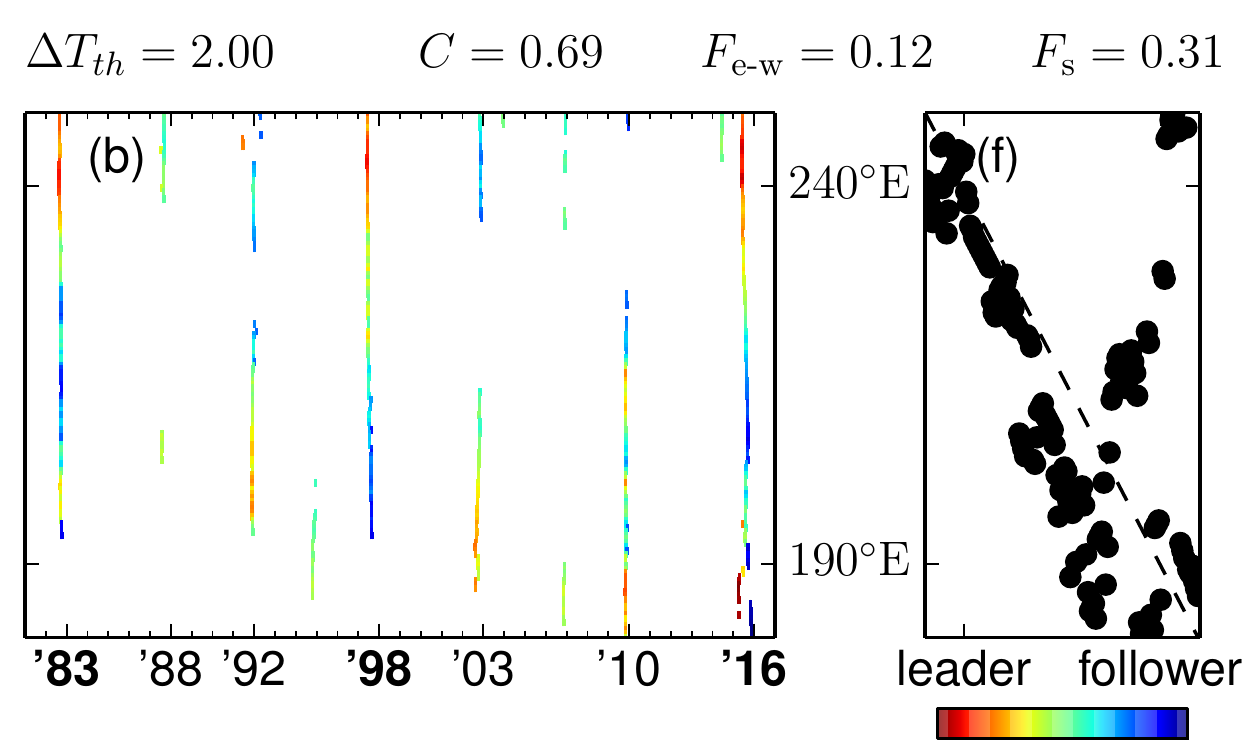}\\[.5em]
 \includegraphics[width=0.4\textwidth]{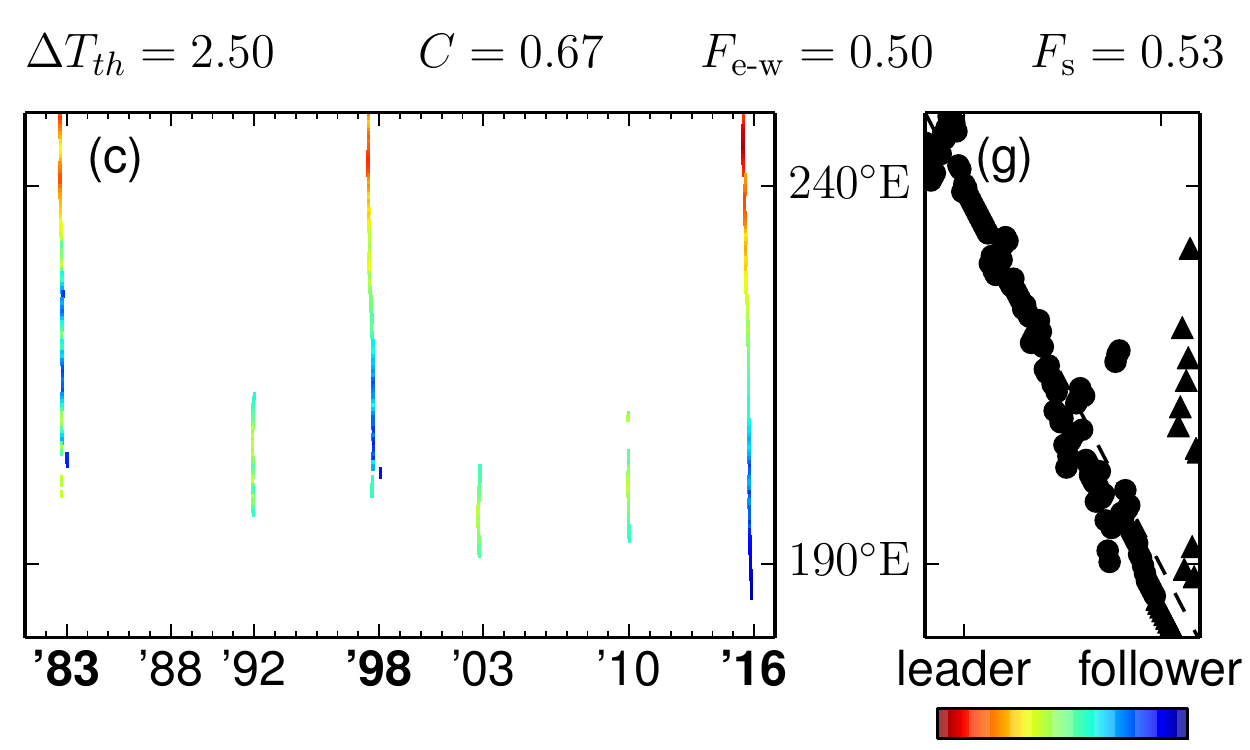}\\[.5em]
 \includegraphics[width=0.4\textwidth]{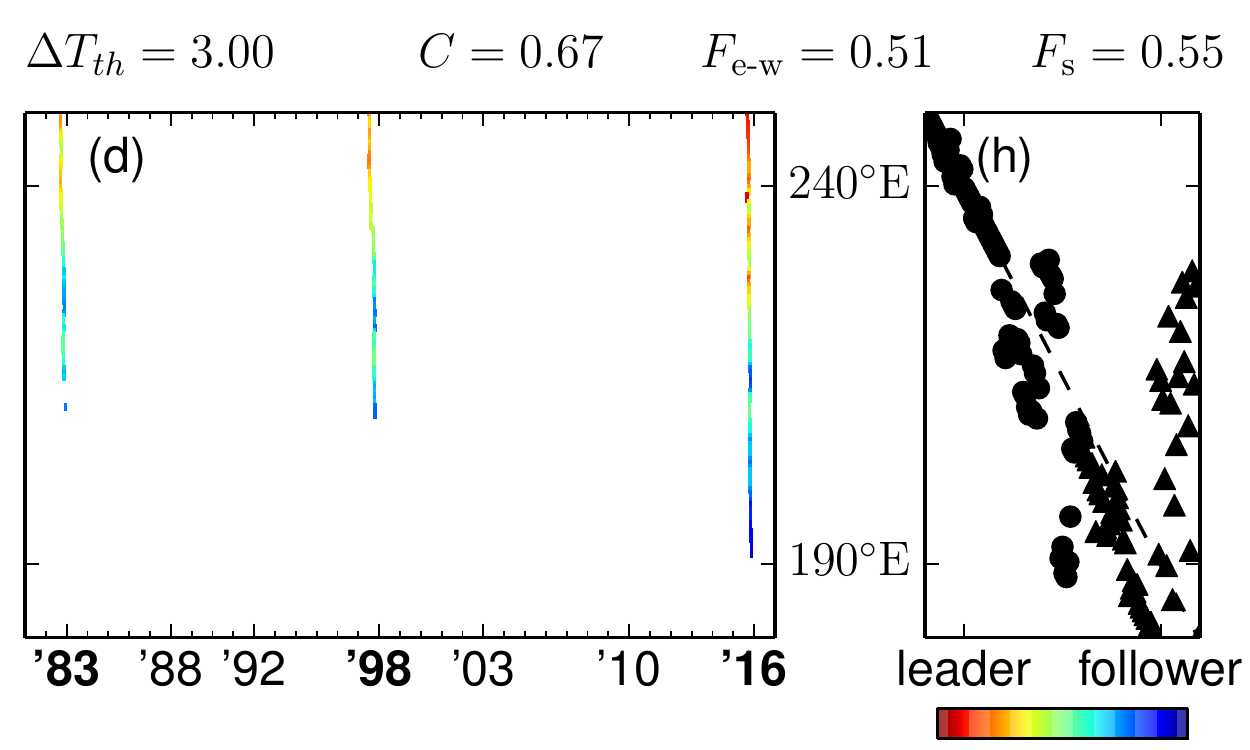}
\caption{Spike Train Order analysis for propagation pattern of daily sea surface 
temperature anomalies within the equatorial strip ($\degS{0.5}\dots\degN{0.5}$, 
$\degE{170}\dots\degE{250}$) smoothed in time using a Gaussian kernel with width 
of $2\,\text{weeks}$. Events are defined as initial threshold crossings of the 
smoothed temperature time series (see text).
From top to bottom four different threshold values have been used: 
$\threshold=\degC{1.5}\dots\degC{3.0}$.
The left graphs (a-d) show the event raster plots with with SPIKE-Order value in 
color coding using an east-to-west ordering.
The years shown on the $x$-axis (representing January 1st of the indicated year) 
denote the moderate and strong (bold face) \elnino\ events.
On the right (e-h), the optimal leader-follower order compared to an initial 
east-to-west sorting is shown (circles: spike trains with at least 2 spikes, 
triangles: spike trains with less than two spikes).
Additionally, the SPIKE-Synchronization $C$ as well as the Synfire 
Indicator for the original (east-to-west) $\Few$, and the optimal sorting 
$\Fopt$ are reported.
}
\label{fig:SPIKE_Order_analysis}
\end{figure}

Starting from the smoothed SST profiles, we performed an event detection by 
identifying the onset of the SST anomaly connected with an \elnino\ in terms of 
an \emph{upward crossing} of the smoothed time series with a temperature anomaly 
threshold $\threshold$.
As we are interested in the initial propagation front for each \elnino\ event, 
we additionally introduce an artificial \emph{refractory period} of 11 months, 
which means threshold crossings occurring within 11 months after previous event 
are disregarded.
\figref{time_series} depicts this procedure for a threshold value 
$\threshold=\degC{2.5}$ leading to the identification of three events in this 
example (vertical lines in \figref{time_series}).
From this, we finally arrive at $N=140$ discrete event series corresponding to 
the onsets of \elnino\ SST anomaly elevations at the different longitudinal 
locations along the dashed line in \figref{el_nino_region}.
The resulting spike trains are depicted in \figref{SPIKE_Order_analysis} on the 
left for four different threshold values $\threshold=\degC{1.5}\dots\degC{3.0}$ 
(step size of $0.5$).
The first thing to note is that this event detection procedure 
is able to correctly identify the \elnino\ occurrences as seen from the 
horizontal event structure in accordance with the \elnino\ years (labeled on the 
$x$-axis in the raster plots a-d) as well as the consistently high 
SPIKE-Synchronization values $C$ for all threshold values.
With increasing temperature threshold, however, only the three strongest 
\elnino\ events are being captured (1983, 1998 and 2016, bold labels in 
\figref{SPIKE_Order_analysis}) as seen from \figref{SPIKE_Order_analysis}c and
\figref{SPIKE_Order_analysis}d.

To quantify the consistency of the propagation patterns, we perform a Spike
Train Order analysis on these event sequences.
The results are also presented in \figref{SPIKE_Order_analysis}.
As a central observation, we found that for the three strong \elnino\ events 
(1983, 1998, 2016), there is a clear consistent propagation of the SST anomaly 
elevation at the equator from east to west as confirmed by a Synfire Indicator 
of $\Few=0.5$ for the east-to-west sorting ($\threshold=\degC{2.5}$,
Fig. \figref{SPIKE_Order_analysis}c).
Furthermore, optimizing the Synfire Indicator by changing the sorting only leads 
to minor improvements ($\Fopt=0.53$) and the resulting optimal order is very 
close to the original east-to-west sorting indicated in
\figref{SPIKE_Order_analysis}g.
Finally, this observation is also confirmed by the SPIKE-Order based color 
coding in the raster plots (c) of the spike trains.
For each of the three major \elnino\ events (1983, 1998, 2016) we see a clear 
tendency of leader (red) to follower (blue) going from east (top) to west 
(bottom) for all three threshold values.
Further increase of the threshold temperature (\figref{SPIKE_Order_analysis}d and h) leads to 
essentially the same observation, but parts of the longitudinal propagation 
might be missed as the SST anomaly maxima stay below threshold.

Moderate \elnino\ events (1992, 2003, 2010), on the other hand, seem to 
propagate in eastwards direction, as seen by opposite directed coloring for those 
years in \figref{SPIKE_Order_analysis}b ($\threshold=2.0$).
Consequently, the Synfire Indicator for the east-to-west sorting is very 
low, $\Few=0.12$, as some of the propagation events are moving in the opposite 
direction which reduces the Synfire Indicator.
Therefore, it is also not possible to find a correct ordering, as indicated by a 
rather low optimized Synfire Indicator of $\Fopt=0.31$ in this case.
This is even more pronounced for the smallest threshold temperature 
$\threshold=1.5$, shown in \figref{SPIKE_Order_analysis}b.
There, essentially all moderate \elnino s are captured and most of those exhibit 
an eastwards propagation, opposed to the strong \elnino s that propagate 
westwards.
Therefore, the initial Synfire Indicator for the east-to-west sorting is 
negative, $\Few=-0.11$ and the optimal sorting shows the sign of a west-to-east 
propagation (\figref{SPIKE_Order_analysis}e) with $\Fopt=0.38$.

\begin{figure}[t]
 \centering
 \includegraphics[width=0.42\textwidth]{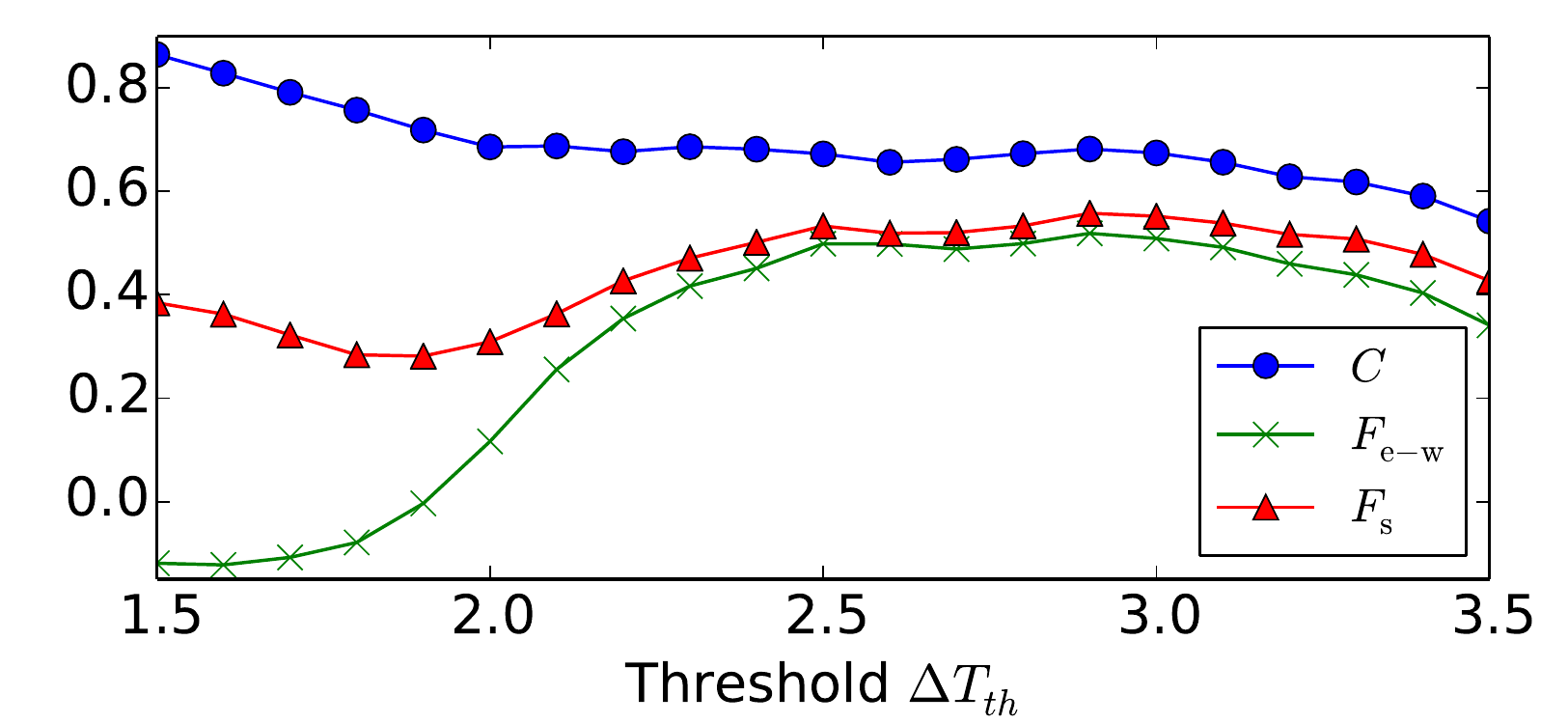}
 \caption{SPIKE-Synchronization $C$ and Synfire Indicator for the 
original sorting (east-to-west) $\Few$ and the optimal sorting $\Fopt$ in 
dependence of the temperature threshold $\threshold$.}
 \label{fig:SPIKE_Order_threshold}
\end{figure}

To better understand the dependency of the Spike Train Order analysis on the 
temperature threshold, we computed the Synfire Indicator for the initial 
east-to-west ordering as well as the value for optimal ordering $\Fopt$ for 
varying thresholds $\threshold=1.5\dots 3.5$ (\figref{SPIKE_Order_threshold}).
Additionally, \figref{SPIKE_Order_threshold} also shows the 
SPIKE-Synchronization values for these parameters.
One first observes that the SPIKE-Synchronization remains rather constant over 
the whole range of threshold values, indicating a consistently good 
identification of the \elnino\ events.
The Synfire Indicator for the original east-to-west sorting $\Few$ on the other 
hand shows a clear increase saturating in a plateau for a threshold value of 
around $\threshold=\degC{2.5}$, where it is then also very close to the optimal 
value $\Fopt$.
This is easily understood by the fact that at temperature thresholds above 
$\degC{2.5}$, only the three strong \elnino s are captured that show a 
consistent westwards propagation, while for values below also weaker \elnino s 
that exhibit eastward propagation enter the analysis.

In summary, we showed that the newly proposed Spike Train Order method is also 
readily applicable to climate data and is able to identify propagation patterns 
in recurring events such as the \elnino.
Our results indicate that close to the equator, strong \elnino\ events exhibit a 
clear westward propagation of an SST front, while for moderate \elnino s we 
found indications for an eastward front propagation.
Interestingly, previous results based on zonal current velocity within the
\elnino\ 3.4 region found eastward propagation for the extreme \elnino\ events
in 1983 and 1998~\citep{Santoso13}, while here we found west-ward propagation
(\figref{SPIKE_Order_analysis}).
This discrepancy is most likely due to the different regions (\elnino\ 3.4 vs
equatorial strip) and methodology (average warming/cooling rates vs front
propagation).



%
%
\section{\label{s:Discussion} Discussion}


Over the last years a wide variety of measures to quantify the synchrony between 
spike trains have been introduced.
Three recent proposals, \textit{ISI-distance} \citep{Kreuz07b, Kreuz09},
\textit{SPIKE-distance} \citep{Kreuz11, Kreuz13}, and
\textit{SPIKE-Synchronization} \citep{Kreuz15, Mulansky15}, share the
desirable property of being time-resolved and parameter-free (time-scale
independent).
However, their bivariate versions are symmetric and in consequence their 
multivariate versions are invariant to changes in the order of spike trains.
None of these measures is designed to provide information about the 
directionality of the propagation patterns.

In the present study we address this issue.
First we use an adaptive coincidence detection as match maker in order to 
identify pairs of coincident spikes.
Then we define two measures, the asymmetric \textit{SPIKE-Order} $D$ and the
symmetric \textit{Spike Train Order} $E$, \highlight{which are particularly useful in
a bivariate representation (pairwise matrix) and as a time-resolved multivariate 
profile, respectively.}
From these two measures we can derive the \textit{Synfire Indicator} $F$, \highlight{a
condensed scalar value that quantifies the overall consistency of the spatio-temporal
propagation patterns in a rigorous and automated way.}
Its maximization allows to sort multiple spike trains from leader to follower.
This is meant purely in the sense of temporal sequence of the spikes.
The question asked is:
\highlight{For which spike trains do spikes tend to occur first and for which do they tend
to occur last?}
We use simulated annealing to search among all permutations of spike trains for
the sorting that resembles as closely as possible a synfire pattern, a perfectly
consistent repetition of the same global propagation pattern.
In a final step we evaluate the statistical significance of the optimized
permutation using a set of carefully constructed spike train surrogates.

We first illustrate the merits of our new approach using artificially generated 
datasets and then apply it to real datasets from two very different fields of 
research, neuroscience and climatology.
In the neurophysiological dataset we analyze Giant Depolarized Potentials 
(GDPs) recorded from mice brain slices in order to search for potential hub 
neurons \citep{Bonifazi09}, whereas in the climate application we quantify the 
consistency of the longitudinal movement of the propagation front of El Ni\~{n}o 
events \citep{Santoso13}.


The new algorithm is conceptually simple, of low computational cost and comes
with an intuitive and straightforward visualization, \highlight{including a
color-coded rasterplot}.
It \highlight{substantially} improves on all the bivariate functionalities
\highlight{of its predecessor directional measure \textit{delay asymmetry}}
(no need for sampled profiles, more intuitive normalization etc.) and could
thus also be used in the context of the pairwise matrices, both normalized or
cumulative, used in complex network theory \citep{Malik10, Boers14}.
However, one of the main advantages of the new algorithm is its multivariate 
nature which opens up completely new kinds of application such as spike train
sorting.

\highlight{One important advantage that the method shares with other techniques
of spike train analysis} is the high flexibility in the definition of events.
For example, when looking at the synchronization of neuronal bursts instead of
individual spikes one can define the events as the onset, the center or the
offset of the activity (e.g., the first, the middle or the last spike of each
burst).
In cases in which a burst of spikes is considered to be equivalent to a single
spike one could introduce some kind of meta-events and then look at
coincidences between these meta-events.

The application of our measures is also not restricted to truly discrete data.
Continuously sampled data can be reduced to a spike train where the only 
information maintained is the timing of the individual events.
Often these event times are obtained in a manner similar to how the neuronal 
spike times are extracted from recordings of neuronal membrane potentials 
(usually done via some kind of thresholding).
Examples of sampled data to which measures of spike train synchrony have been 
applied include EEG data \citep{QuianQuiroga02b, Kreuz13, Rosario15} and, 
outside of neuroscience, stock market velocity \citep{Zhou03} -  and rainfall 
events \citep{Malik10}.

\highlight{The algorithm is particularly suited for datasets with a high value of
SPIKE-Synchronization.
According to the coincidence criterion (Eq. \ref{eq:Coincidence-MaxDist}) these are
spike trains that include sequences of global events for which the interval
between successive events is at least twice as large as the propagation time within
an event.
For these datasets the Synfire Indicator evaluates the consistency of the order
within these well separated global events.}
The universality of the phenomenon, repetitive propagation patterns, makes our 
new algorithm applicable in a wide array of fields such as medical sciences,
seismology, oceanography, meteorology or climatology.
For example, the duration of an epileptic seizure is typically much shorter than 
the interval between two successive seizures.
Also the time it takes a storm front to cross a specific region is typically 
much smaller than the time to the next storm.
Many other repetitive propagation phenomena exhibit similar ratios of 
characteristic time scales.


In order to understand the scope of our proposed algorithm it is important to 
understand what it is not designed to achieve.
The method deals purely with relative order, it does not consider the length
of absolute delays.
\highlight{Moreover, while the instantaneous coincidence criterion makes the method
time-scale independent, parameter-free and easy to use, it also renders it
insensitive to patterns involving spikes that are not immediately adjacent.}
Many other, \highlight{typically more complicated,} methods have been designed to
characterize the detailed spatio-temporal structure in large neuronal networks
\highlight{\citep{Schneider06, Gansel12}} or to detect hierarchically structured
spike-train communities \highlight{\citep{Billeh14, Russo17}}.
The method is also not designed to detect neuronal synfire chains (in the strict 
sense of the word) in massively parallel data.
For this task other statistical methods based on some forms of pattern detection 
have been developed \citep{Schrader08, Gerstein12}.

Another caveat concerns causality.
While a significant value of the Synfire Indicator $F_s$ in our algorithm 
clearly shows the presence of a preferred temporal order of some signals with 
respect to others, it does not necessarily prove a driver-responder 
relationship.
There are other methods that have been developed for this kind of system 
dynamics analysis (e.g., \cite{Andrzejak11}).
But even for such methods causality is always a strong claim.
In fact, the two signals might be driven by a common hidden source and a 
consistent leader (follower) could just indicate a drive with a smaller (larger) 
delay.
Similarly, internal delay loops in one of the two systems can also fool the 
interpretation.


There are a number of possible directions for future research, both from a 
methods and from a data point of view.
Regarding the algorithm, for the coincidence detection it would be straightforward 
to limit the range of allowed time lags by incorporating information about the 
expected speed of propagation \citep{Malik10}.
One could introduce a minimum time lag in order to ensure causality and/or
limit the maximally allowed time lag in order to focus on meaningful
propagation of activity.
In principle the range of allowed time lags could even be selected individually 
for each pair of spike trains depending on the known properties of the 
connectivity between the respective two neurons.
Importantly, even with such type of time lag restrictions in place, it has still 
to be guaranteed that each spike can be part of at most one coincidence.

A follow-up task for our neurophysiological data would be to investigate to what 
extent the neurons that are identified as leading by our analysis are identical 
to the so-called hub neurons \citep{Bonifazi09}, i.e. neurons with a much higher 
than average degree of connectivity within the network.
Regarding the \elnino\ analysis, the difference in propagation directions
observed for the strong and weak \elnino\ events remains an open question.
Furthermore, expanding the analysis to wider regions, i.e.\
$\degS{5}\dots\degN{5}$ and verifying the consistency of the observed
propagation patterns is an interesting topic for future research.
Finally, the relation to results based on average warming/cooling
rates~\citep{Santoso13} requires further investigations.

The algorithm will be readily applicable for everyone since, together with the
existing symmetric measures ISI-distance, SPIKE-distance, and
SPIKE-Synchronization, SPIKE-Order is implemented in in three publicly available
software packages, the Matlab-based graphical user interface
SPIKY~\footnote{http://www.fi.isc.cnr.it/users/thomas.kreuz/Source-Code/SPIKY.html}
\citep{Kreuz15},
cSPIKE~\footnote{http://www.fi.isc.cnr.it/users/thomas.kreuz/Source-Code/cSPIKE.html}
(Matlab command line with MEX-files), and the Python library
PySpike~\footnote{http://www.pyspike.de} \citep{Mulansky16}. 

%
%

\begin{appendix} \label{s:Appendix}

\section{\label{App-s:HM-Data} Neurophysiological dataset}

The neurophysiological data analyzed in Section \ref{ss:Results-Neuro} were 
recorded via fast multicellular calcium imaging in acute CA3 hippocampal slices
from juvenile mice.
\highlight{The CA3 region has a strong recurrent excitatory connectivity \citep{Amaral95}.
This distinct feature is suggested to be crucial for memory encoding and pattern
completion and thus memory retrieval \citep{Leutgeb07}.
During memory retrieval in rodents, population bursts of the CA3 lead to high
frequency stimulation of the efferent regions, so called sharp wave ripples
\citep{Chrobak96}. 
In the juvenile hippocampus, due to a higher chloride reversal potential
in the CA3 pyramidal cells, the GABA-ergic system is excitatory \citep{BenAri89}.
GABA-ergic interneurons have been shown to serve as so called hub neurons that
trigger the GDPs \citep{Bonifazi09}.}

The recordings were performed by the group of Heinz Beck at the Department
of Epileptology, University of Bonn, Germany, prior to and independently from
the design of  this study.
Transversal acute brain slices ($300 \mu m$ thick) were prepared from $5$ to 
$10$-day-old (P5-P10) C57BL/6 mice (Charles River, n = $19$ slices).
Slice preparation, calcium imaging and data analysis were performed as
previously described in \citep{Hedrich14}.
For AM-loading of brain slices with OGB1-AM we used a protocol modified from 
\citep{Crepel07}.
Multicellular calcium imaging was done using a homemade single planar
illumination microscope (SPIM) modified from
\citep{Holekamp08}.
Movies were recorded at a frame rate of $200$ Hz over a minimal length of $5$
min up to $30$ min to record a sufficient amount of spontaneous activity.
Time points of cell activity from the imaging data were defined as the onsets
of Ca$^2$ events in fluorescence traces of all individual cells using the
maximum of the second derivative of each event \citep{Henze00}.

In order to test the Spike Train Order algorithm, datasets were chosen that
exhibited at least three global GDPs during the recording ($n = 5$).
For one dataset the surrogate analysis described in Section
\ref{ss:Statistical-Significance} proved to be unfeasible due to its excessive
density of spontaneous activity.
Therefore this dataset was discarded from further analysis, so the final number
of datasets analyzed was $n = 4$.

\section{\label{App-s:ElNino-Data} El Ni\~{n}o dataset}

The data analyzed in Section \ref{ss:Results-Climate} describe one of the most 
well-known global climate effects, the so called \elnino, the warm phase of the 
\elnino\ Southern Oscillation (ENSO).
It is usually characterized by an increased sea surface temperature (SST) in the 
central and eastern tropical Pacific Ocean typically during the months
September -- February~\citep{Trenberth97}.
According to the U.S.\ National Oceanic and Atmospheric Administration (NOAA), 
an \elnino\ event is identified by an increased three-month moving average SST 
by $\SI{0.5}{\degreeCelsius}$ over at least six months.
\elnino s can last from nine months up to two years and typically occur in 
irregular intervals of two to seven years.

The analysis presented in Section \ref{ss:Results-Climate} uses the Optimum 
Interpolated Sea Surface Temperature (OISST) data provided by NOAA.
These data result from a high-resolution blended analysis (spatial resolution of 
$\SI{0.25}{\degree}$) of daily SST and ice constructed by combining observations    
from different platforms (satellites, ships, buoys) on a regular global grid 
with a time range from 1981 to 2016~\citep{Reynolds07}.
The daily SST data is centralized by the long-term daily mean resulting in daily 
SST anomaly data (deviations from long-term mean) that form the basis of this 
analysis.


The area used to define \elnino\ events, the \elnino\ 3.4 region, stretches from 
$\degS{5}\dots\degN{5}$ in latitude and $\degE{190}\dots\degE{240}$ in 
longitude, as shown in \figref{el_nino_region}.
However, for the latitudinal direction we here focused only on a small central
strip around the equator, i.e.\ from $\degS{0.5}\dots\degN{0.5}$ over a
longitudinal stretch consistent with the \elnino\ region, i.e.\ from
$\degE{180}\dots\degE{250}$ (dashed line in \figref{el_nino_region}).
Note, that this focus on the small region around the equator is the main 
difference to most previous works that studied the propagation of SST anomalies 
averaged over a much larger latitudinal extent, e.g.\ $\degS{5}\dots\degN{5}$ in 
\cite{Santoso13,Antico16}.
In latitudinal direction, we average over the whole strip (four grid points), 
while in longitudinal direction an averaging over two grid points is performed 
resulting in $\SI{0.5}{\degree}$ of spatial resolution and a total of $N=140$ 
time series of daily SST anomalies.
We disregard short-term fluctuation by applying a Gaussian smoothing with width 
$\sigma=\SI{14} {days}$.
In \figref{time_series} we show the time series resulting from this 
procedure exemplarily for the center of the observed region, $\degE{215}$.

For the Sea Surface Temperature data we use a threshold criterion to identify 
the moving high temperature fronts of the El Ni\~{n}o events.
The threshold determines the signal-to-noise ratio of the data: for small 
values even weaker events have an effect on the result, whereas for higher 
values the analysis is focused on the strongest El Ni\~{n}o events only.
Due to the variability of the propagation patterns it might happen that in 
successive years the threshold is surpassed in different regions.
However, since the aim of the analysis is to look at the propagation of 
individual (seasonal) fronts we suppress coincidences between threshold 
crossings from different years.
We still use the adaptive coincidence detection from Eq. 
\ref{eq:Coincidence-MaxDist} but define a maximum coincidence window 
$\tau_{max}$ which in this case is set to $9$ months.

\end{appendix}

\vspace{0.7cm}

\begin{thanks}
\section{\label{s:Acknowledgement} \textbf{Acknowledgements}}
We are indebted to Heinz Beck from the Department of Epileptology, University
of Bonn, Germany, for providing the neurophysiological data.
We thank Markus Abel, Roman Bauer, Heinz Beck, Nebojsa Bozanic, Christoph
Gebhardt, Marcus Kaiser, Stefano Luccioli, Florian Mormann, Giorgia Paratore,
and Alessandro Torcini for useful discussions.
Finally, we are grateful to Irene Malvestio and Bernd Mulansky for carefully
reading the manuscript.
TK thanks Marcus Kaiser and his group for hosting him at the University of 
Newcastle, UK.

We acknowledge funding support from the European Commission through the Marie 
Curie Initial Training Network `Neural Engineering Transformative Technologies 
(NETT)', project 289146 (TK and MM), and through the European Joint Doctorate 
`Complex oscillatory systems: Modeling and Analysis (COSMOS)', project 642563 
(TK and ES).
TK also acknowledges the Italian Ministry of Foreign Affairs regarding the
activity of the Joint Italian-Israeli Laboratory on Neuroscience.
\end{thanks}


\bibliographystyle{elsart-harv}

\end{document}